# Making sense of AI systems development

Mateusz Dolata and Kevin Crowston

**Abstract** — We identify and describe episodes of sensemaking around challenges in modern Artificial-Intelligence (AI)-based systems development that emerged in projects carried out by IBM and client companies. All projects used IBM Watson as the development platform for building tailored AI-based solutions to support workers or customers of the client companies. Yet, many of the projects turned out to be significantly more challenging than IBM and its clients had expected. The analysis reveals that project members struggled to establish reliable meanings about the technology, the project, context, and data to act upon. The project members report multiple aspects of the projects that they were not expecting to need to make sense of yet were problematic. Many issues bear upon the current-generation AI's inherent characteristics, such as dependency on large data sets and continuous improvement as more data becomes available. Those characteristics increase the complexity of the projects and call for balanced mindfulness to avoid unexpected problems.

**Index Terms**— Artificial Intelligence, Empirical study, Industry, Social issues, Software engineering, Systems development

## INTRODUCTION

THE potential of Artificial Intelligence (AI) technology has led to its increased use by companies and individuals [1], [2], [3]. This increase has spurred the development of tailored solutions using AI tools and data sets provided by the user companies, currently further propelled by the rise of large language models (LLMs). However, the distinctive properties of modern AI technologies pose numerous challenges to system development [4], [5].

According to US National Science and Technology Council, "AI enables computers and other automated systems to perform tasks that have historically required human cognition and human decision-making abilities" [6]. The term encompasses diverse technologies including natural language processing, machine learning, robotic process automation, chatbots, information retrieval, hypothesis generation, and image processing and others [7].

Minsky [8] distinguished between top-down and bottom-up approaches to simulating human cognition. Top-down approaches use logic and/or symbolic rules to represent human reasoning, while bottom-up approaches capture reasoning in complex structures derived from data, e.g., neural networks. Minsky described top-down systems as logical, symbolic, and neat, and bottom-up systems as analogical, connectionist, and scruffy. Top-down systems are deterministic and rule-based, while bottom-up ones are probabilistic and statistics-based [4]. Historically, SE has focused on deterministic systems prevalent in practice [7], but advances in processing power and data availability have led to the recent focus on probabilistic systems for AI.

TABLE 1
COMPARISON BETWEEN CONVENTIONAL SOFTWARE DEVELOPMENT AND AI-BASED DEVELOPMENT USING PROBABILISTIC REASONING

| | Conventional / Legacy Software | State-of-the-Art AI |
|---|---|---|
| Features | Logical | Analogical |
| | Symbolic | Connectionist |
| | Neat | Scruffy |
| | Deterministic and Rule-based | Probabilistic and Statistics-based |
| Implications | Divide, Conquer, and Merge ⇢ Solution | Accumulate, Train, and Test ⇢ Learning |
| | Decision performance grows with the invested effort | Decision performance is a function of data, data processing, and algorithm fit |
| | Functionality is a sum of functionalities of the components | Functionality emerges from the interaction between components |
| | Upgrade ≈ new functionality | Upgrade ≈ higher accuracy |

The inherent differences between these approaches have implications for development practice [4], [9], [10]. We highlight four aspects. First, top-down systems can be developed using a divide-and-conquer strategy that breaks down logical reasoning chains into individual rules or functions. However, probabilistic models are essentially huge collections of seemingly random rules, blocking insight into the model's workings and obstructing separation of capabilities. As a result, bottom-up development comprises cycles of data accumulation, training, and repeated accuracy testing of the model as a whole. We suggest the term "accumulate-train-and-test" for this strategy.

Second, performance of conventional top-down systems is expected to improve with invested effort, as work focuses on enhancing functions and reducing developers' errors in specifying rules. However, the quality of probabilistic systems, built bottom-up and rooted in data, relies on the data and algorithms used. Extra optimization efforts, such as sourcing more data or refining pre-processing, might improve but possibly instead worsen the output. In other words, it's uncertain if and how additional effort will impact system performance.

Third, conventional top-down systems, with hierarchically organized subsystems, are complicated but understandable. Probabilistic systems, however, depend on the non-deterministic interaction between data and models, making them *complex* systems [11], [12]. Complicated and complex systems differ: "A complicated system is one that can be described in terms of its individual constituents, whereas a complex system is one that can be described only in terms of the interactions among the constituents" [13, p. 67]. One cannot assess, fully describe, or understand a complex system by examining its constituents in isolation. It is subject to nonlinear relationships, spontaneous (re)orderings, adaptations, and highly dynamic interactions. Neither its internal configuration (i.e., the

- *Mateusz Dolata is with the Department of Informatics, University of Zurich, Zurich, Binzmuehlestrasse 14. E-mail: dolata@ifi.uzh.ch.*
- *Kevin Crowston is with the School of Information Systems, Syracuse University, Syracuse, Hinds Hall 230. E-mail: crowston@syr.edu.*

interactions between its components) nor the relationship with the environment can be considered stable [11].

The final implication concerns system upgrades. Conventional top-down systems are upgraded through addition of new features, corrected rules, and extended symbols, while probabilistic systems add enhancing accuracy to benefit existing and potentially new use cases. Table 1 summarizes these features and implications we deduced from them. Because of these differences, current AI-based system development is fundamentally different from deterministic systems, challenging conventional software development knowledge, leading to possible unexpected and problematic outcomes during development.

A further complication to AI-based development in the current environment is that, due to AI's technological demands and reliance on processing large data sets, many companies lack the computing infrastructure to build their own AI systems. As well, training and maintaining an AI ecosystem and ML models is costly and risky for many organizations, as they may fall behind larger AI providers in this dynamic market. Generative language models trained a year or two ago may underperform compared to recent LLMs like GPT or PaLM. To address these concerns, many IT providers offer AI resources via software development platforms (DPs), providing modules, digital infrastructure, standards, and services for new applications [14], [15], forming what we call *AI-based development platforms* (AIDPs). AIDPs provide modules such as text and image processing, speech recognition, data extraction, and natural language understanding to be integrated with client data and applications. Modules can be updated as the technology improves. AIDP providers also supply the necessary hardware, software, and runtime environments for deploying or training new ML models. AIDPs have evolved into ecosystems incorporating elements from other providers, such as data hosting services [16]. They are thus "foundations upon which other firms can build complementary products, services or technologies" [17, p. 54]. Several AIDPs, including IBM Watson, Google Cloud AI, Microsoft Azure ML Studio, and Keras.io, are available. The current LLM trend shows that AIDPs continue to be relevant. In the remainder of this paper, we focus on systems built on these platforms.

In addition to technical gaps, companies often also lack specialized personnel to keep up with rapid AI advancements, leading them to seek external support [18]. Smaller companies may outsource to freelancers or agencies for short-term collaborations [16], while larger organizations often form long-term partnerships with bigger providers for continuity and security. Similar trends were observed in the ERP software market [19], [20]. Platform providers in particular offer support to developers and companies. For instance, IBM provides industry-specific consulting for Watson, while Microsoft collaborates with partner firms for Azure Machine Learning consultation. Numerous organizations are partnering with AIDP providers to leverage AI technologies [21]. As of late 2022, around 30,000[1] organizations, including Fortune 500[2] companies, were estimated to use AIDPs. With the recent surge in LLMs, this number is likely higher. We use the term *AIDP-based development* to highlight the reliance on external resources, toolkits, hardware, documentation, release processes, and platform-specific skills from partners. Understanding how this AI delivery method affects developers and development processes is crucial to making it reliably successful.

An emerging research stream on Software Engineering for Artificial Intelligence (SE4AI) has delved into issues related to AI development [9], [22]. SE4AI differs from traditional SE, posing technical and organizational challenges [4], [23], [24]. These differences will likely change the socio-cognitive SE processes. In this paper we focus in particular on challenges to individual and collective sensemaking [25], [26], [27], applying a sensemaking perspective to identify the response to issues during AIDP projects.

Following this approach, we identify frames that shape stakeholders' expectations, cues that cause breakdowns of those frames, and new meanings that emerge through collective action. By frame, we mean a set of initial assumptions and interpretations about an event or phenomenon that guide action. Breakdowns happen if insights from the context are seen to invalidate those assumptions and meanings. For instance, in the context of AIDP projects, a frame might be the expected division of labor between a vendor and the client company. However, during the project new tasks might emerge (a cue) for which responsibility is unclear (a breakdown in the frame) that require rethinking and negotiating who is responsible for which tasks (new meaning). Past research assumes that sensemaking is an ongoing process which mostly happens implicitly, but suggests that paying attention to sensemaking processes is beneficial as it raises sensitivity to the possibility of problems and the need to respond [25].

We find that AIDPs amplify project complexity, resulting in more challenging emergent behavior. The complexity stems from features like probabilistic processing, opaqueness, and reliance on new or pretrained models, making progress hard to assess. AIDP projects also introduce tasks beyond conventional SE, including data collection and selection, ML model training, and accuracy evaluation. These tasks create new roles and responsibilities that, as our data indicates, shift unpredictably due to technology, data, or context demands. AIDP projects are thus more complex for both providers and clients, making participants' assumptions and heuristics less reliable. Clients and vendors engage in additional rounds of sensemaking of the technology and development efforts, leading to new project perceptions. The new meanings accommodate the element of unexpected discoveries and fluctuations. Our study outlines these projects' complexities, highlighting the importance of inadequate expectations. It identifies four sensemaking areas for AIDP projects: data, technology / platform, project, and context. The multi-target sensemaking is crucial for progress in AIDP projects.

This research contributes to the ongoing SE4AI discourse in several ways. First, it complements existing analyses of sensemaking in AI system development [16], [18], [28], covering all potential areas instead of focusing only on project context [18] or ML models [28]. Second, it advances the conversation about AI's impact on SE processes [4], [23], [24] by focusing on collaborative, inter-organizational projects and the socio-cognitive aspects of software development. Lastly, it attends to a specific yet popular mode of AI delivery, AIDPs. While this may limit generalizability, it offers deep insight into a coherent set of projects, reducing abstraction risks. Overall, the paper studies how AI's and AIDPs' features impact the collaborative engineering process between providers and clients.

[1] enlyft.com / tech / machine-learning; retrieved on November 9, 2022.

[2] www.appsruntheworld.com / customers-database / products / view / microsoft-azure-ai-platform or ~ / ibm-watson; retrieved on Nov 9, 2022.

## RELATED WORK

SE4AI is a growing research area that emphasizes the unique nature of AI-based development compared to traditional software projects and suggests potential solutions to these challenges. Key sources of uncertainty, such as data and the probabilistic nature of Machine Learning, create numerous challenges for developers, as highlighted in various studies [4], [9], [16], [18], [23], [24], [29], [30]. Recent reviews categorize these challenges [9], [10], [22], [31]. For instance, Martínez-Fernández et al. [9] and Giray [22] provide comprehensive overviews, identifying more than 94 unique challenges classified based on the SWEBOK Knowledge Areas [32]. However, they stress that SWEBOK needs extension to consider increased data and training dependency of ML-based approaches, and the fact that their runtime behavior might change over time due to incremental training or drift of input data patterns. Other meta-studies note that AI poses additional specific, previously unknown challenges. For instance, the lack of explainability of most ML approaches negatively impacts SE professional practice, e.g., not understanding why a ML model produces the output it does can be hard to explain to customers [33], [34]. In the following we discuss the challenges following the organization proposed by Martínez-Fernández et al. [9].

The largest cluster of challenges pertains to Software Testing and Quality [9]. The absence of test data or test cases hampers the systematic evaluation of AI-based systems [35]. Finding appropriate measures to assess system output is a challenge too [36]. Many statistical metrics do not translate easily into business requirements. Further, many AI systems produce outputs that cannot be easily classified as right or wrong. Instead, output quality depends on user assessment, which may consider factors like usefulness or correctness. This assessment might be costly, limiting access to adequate test data. Proposed solutions involve using AI for testing [36]. However, this approach shifts responsibility from users or clients to developers, conflicting with domain knowledge distribution.

The second cluster relates to Software Requirements Engineering [9]. The process of eliciting and specifying requirements is a key topic in SE literature, as it involves translating a vague organizational context into a systematic technical description [37], [38], [39], [40]. This translation creates challenges in professional communication among stakeholders to deal with uncertainty and ambiguities. AI can exacerbate these issues, as stakeholders often expect 100% accuracy and clarity, or may fear systematic biases or AI in general [9]. The unpredictability of results also obscures validation of requirements fulfillment [40]. While there is a recognized need for AI requirements engineering approaches, their development is slow [41], [42].

The third cluster, about SE Models[3], Methods, and Processes, finds that these elements currently lack sufficient support for practitioners. They fail to account for costs and activities associated with data pre-processing, labelling, management, and the experimental nature of ML model development [43], [44], [45]. The recent adoption of the Machine Learning Operations (MLOps) paradigm provides a structured approach to building and maintaining AI-based software and a meta-level view of temporal order and logical dependencies between activities [46], [47], [48], [49]. Research on ML pipeline architectures also offers an abstract model of AI-based applications [50], [51]. However, applications of these models can themselves create further challenges, especially in an interorganizational context due to differing SE practices, organizational history, software perspectives, and data privacy issues [52].

The fourth cluster relates to software creation, including Design, Construction, and Maintenance [9]. The CACE ("changing anything changes everything") principle, applicable to AI due to its complex (not just complicated) nature, poses a significant difficulty. The functionality of AI systems stems from component interactions, making divide and conquer strategies unsuitable [18]. This issue is amplified by parallel processing for large model training, reliance on external AI frameworks, and difficulties in reproducing errors and unwanted ML component responses [44], [45], [53]. These challenges are due to AI's complexity and inherent characteristics like opaqueness, probabilistic processing, and data dependency. However, the effects of these challenges on development team's socio-cognitive processes, particularly in inter-organizational collaboration, remain unclear.

Further SWEBOK Knowledge Areas are only minimally covered in the primary studies reviewed. Out of the 248 studies considered by Martínez-Fernández et al. [9], only two focus on Software Engineering Management [16], [54], one on Software Configuration Management [55], and one on SE Professional Practice [56]. The lack of studies in these areas is unfortunate, since the technical and practical challenges of AI are likely to affect socio-cognitive processes among developers and stakeholders, creating management issues. Understanding these implications can guide management and increase the success and value of AI-based software projects. In particular, Wolf and Paine [16] suggest that sensemaking is crucial in AI-based projects, particularly in business context, AI/ML environments and AI/ML model ecosystems. However, their study abstracts from project configuration, technical stack, and the process of sensemaking itself (initial beliefs, breakdown sources, revised beliefs [25]). By focusing on AIDP-based projects involving clients and providers, we analyze sensemaking processes and their relation to the discussed challenges.

Overall, while much literature addresses AI software project challenges, there is little on management and professional practice. The unique characteristics of AI and its technical challenges can have varying implications depending on the context, yet studies adopt a developer's perspective, overlooking project configurations such as single-company, or collaborative settings [9]. This lacuna is surprising as there are significant differences between contexts like outsourcing and insourcing [57], [58]. Current literature offers limited guidance, and proposed paradigms like MLOps are still in their early stages, lacking empirical validation [48], [59]. We propose to examine how AI-typical challenges affect socio-cognitive processes in AIDP-based software development and how stakeholders address these issues to provide better guidance for project setup and management.

[3] Martínez-Fernández et al. [9] subsume some challenges related to statistical, ML modeling under SE Models and Methods Knowledge Area. For instance, they mention ML model overfit as a challenge. For us, such aspects are related to Software Design and Construction as they generate the desired functionality and impact software structure and architecture. We refer to those challenges in the subsequent paragraph.

## Theory

The study started with IBM's noticing challenges in Watson-related projects. Specifically, IBM's Swiss sales department had noticed that Watson projects suffered from increased costs and project duration, and fewer than expected follow-up projects with the client companies compared to non-Watson projects. We sought to understand why an experienced development company like IBM working with the experienced IT departments of large client companies ran into problems. While our study offers many potential lessons, we were initially struck by the struggles clients and providers had to make sense of the novel character of the technology and the appropriate client-vendor relationship to manage the project. They described frequent and significant shifts of their perceptions concerning the technology, the project, and the partners. Their concerns called our attention to the issue of sensemaking as a core aspect of system development [16], [60], [61]. Accordingly, our research objective is to *identify and describe problems in making sense of AIDP-based projects due to the distinctive nature of the platform*.

Sensemaking emerged originally as a theoretical lens to study socio-cognitive processes and human action in *complex* situations [25], [26], [62], [63], [64]. Complexity refers to "the lack of a tight linear structure and the high probability of unexpected synergistic (possibly negative) interactions among component parts" [65, p. 276] or a "combination of lack of control and inability to comprehend what is happening" [66, p. 33]. As noted above, modern AI systems are characterized by complexity, as their constituents enter unpredictable interactions with each other. In the face of such complexity, human actors must conduct continuous sensemaking to remain engaged in the action.

Sensemaking was originally proposed as a lens to study events with an unexpected, negative outcome. The original case described by Weick is the Mann Gulch disaster in which 13 firefighters died [27]. In this case, he shows that ignoring contradictory cues, belated sensemaking, and differences in meanings members of the brigade attached to the situation resulted in the tragedy. Current research broadens attention from negative outcomes to identifying problems in cognitive and social processes involved in any project or complex situation [62]. The perspective is appropriate for studying SE, which involves a variety of cognitive and communicative processes. applying this perspective, SE can benefit from a better understanding of how assumptions made by stakeholders shape the development processes, and how they are reflected in solutions developed in those processes. We can develop guidance on how developers should deal with unexpected events in their projects and how they can early spot upcoming difficulties.

In SE, a need for sensemaking can be driven by breakdowns due to vague or incomplete requirements, poor risk management, and buffer time erosion, leading to crisis and fire-fighting behaviors [67]. In inter-organizational projects, the social nature of collaboration, opaqueness regarding internal processes and goals of the participating organizations, and insufficient communication patterns (e.g., in offshore projects) generate further sources for tentative frames, incompatible meanings, and breakdowns [61]. In a complex situation, "a number of parties handling a problem are unable to obtain precisely the same information about the problem so that many interpretations of the problem exist" [68, p. 50]. As a result, observing the multiparty character in AIDP-based projects is central to provide a holistic sensemaking perspective.

*The temporality of sensemaking.* Originally, sensemaking was defined as a backward-oriented process, i.e., the past and currently emerging cues were used to establish tentative meaning and act upon it [27], [69]. Later research has brought up the notion of prospective sensemaking, which considers potential consequences of the actions one is planning to undertake [13], [70]. In prospective sensemaking, people rely on sufficiency and plausibility rather than comprehensiveness and accuracy, which make meanings even more tentative and less reliable than in retrospective sensemaking [25]. Prospective and retrospective sensemaking are intermingled and influence each other.

*The process of sensemaking.* Through explorations of sensemaking, research has documented the sequence of steps involved [13], [71]. Engagement with the environment starts with *expectations* based on past experiences [70], inspirations, frames [63], inputs [25], or preconceptions [69]. Combined with cues from the environment, the expectations form initial meanings about what is likely to happen [25], [64]. As the engagement continues, new cues are incorporated into those meanings for the subject to act upon. However, under time pressure or stress, subjects will act selectively and prioritize cues confirming their initial meanings and discount those that do not align [13].

Eventually though, when the volume of opposing cues grows, causing the environment to become ambiguous, subjects experience a *breakdown*: they face a loss of sense, or the sense becomes increasingly elusive [64]. These experiences trigger efforts at sensemaking through which the subject tries to *resume* the interrupted activity and stay in action. People draw on surrogate frames, e.g., those offered by the organizational context (e.g., plans, constraints, justifications) or society (e.g., generic frames and structures) to resume and sustain the actions [64]. Alternatively, people might recover by identifying new meanings which suggest an alternative course of action [64]. In either case, once an adequate frame for action is reestablished, the subject will enact the new meanings until another breakdown emerges. We use this sequential view on sensemaking to analyze how it unfolds in AIDP projects.

## Methods

*Study setting*. Our study is set in the context of IBM Watson project development in Switzerland. Watson is an AIDP including business-ready tools and solutions designed to improve the adoption of AI techniques in work environments[4]. It emerged by modularization, re-training, and extension of a question-answering engine known for its successful participation in a TV quiz show in 2011 [7]. Shortly thereafter, IBM started projects with other client companies to leverage the abilities in work-related contexts [72], [73]. In 2013, IBM opened the DP for use by independent developers[5] and since then has continuously extended its functionalities, added new APIs, tools, and models, all under the banner of "cognitive computing".

In parallel, IBM engaged in commercial collaborations with organizations from around the world to identify and develop use cases for use of Watson. Those projects address specific needs of clients and rely on the use of the

4 www.ibm.com/watson; retrieved on November 9, 2022.

5 www.forbes.com/sites/bruceupbin/2013/11/14/ibm-opens-up-watson-as-a-web-service/; retrieved on November 9, 2022.

client's data sets. They include consulting and development services. The client pays for the services offered by IBM though hour rates and other agreements are not public. Official statistics about Watson projects are also not public, but IBM refers to over 100 million users benefitting from Watson applications[6]. In these projects, IBM takes the role of a vendor. Accordingly, it provides knowledge and resources to support the client in developing an application. As we started collecting data in March 2017, IBM already had much experience with Watson projects around the world. According to internal information, at the time of our study, IBM Switzerland had about 3 years of experience in running Watson projects and over 50 projects running or recently completed.

*Study design.* This paper follows a qualitative research methodology. The primary mode of insight is retrospective analysis of sensemaking episodes depicted by project members. We adhere to the ideal of exploratory research [74] combined with analysis inspired by critical incidents techniques [75], [76]. We strive to understand patterns related to problems faced in AIDP-based projects and the nature of sensemaking to address these and therefore study multiple cases [77]. The researchers are independent of both IBM and its clients. We rely on data, observer, and theory triangulation to enhance precision and accommodate for a broader picture of the studied phenomenon [78].

*Data elicitation*. Data for the study comes from interviews with informants from IBM and its clients. To select interviewees, two senior IBM managers scanned all IBM Watson projects in Switzerland, resulting in 21 selected projects involving 17 companies located in Switzerland. The projects between IBM and clients combined three goals: yielding an AIDP-based application for use by the client, investigating potentials of long-term business cooperation, and giving the client hands-on experience with AI and Watson. Consequently, many were referred to as exploratory projects. For instance, a major Swiss insurance company envisioned an application that would help its internal underwriting department collect and summarize their own and publicly available information about small businesses to assess their risk levels and provide a more adequate insurance offering. The supplemental material lists all projects and their specific purpose.

Between March and May 2017, our team carried out 36 semi-structured interviews. The interviewees were IBM-side project managers, client-side project managers, and lead IBM consultants or industry solution architects. One person was a client-side developer who represented a project manager. The focus is on high-level professionals as opposed to the perspective of developers covered extensively in previous literature [9]. As the goal was to report on the overall perception of the projects, we aimed for people who could provide that overview rather than trying to represent different project roles.

For 17 projects, we conducted interviews with representatives of the client and IBM. Two companies were involved in two different parallel Watson projects and another company in three: interviewees from those companies reported on all projects in their interviews. Client representatives were not available in the remaining cases, so we only interviewed the IBM side. Three client-side interviewees were women; eleven were men. Five IBM-side interviewees were women, 17 were men. Employees from all organizations reported that they had previous experiences in client-vendor collaborations. All interviewees had at least two years of experience working either for IBM or the client companies, so they knew the context of their work.

To guide the interviews, four main areas of interest and multiple open questions were prepared but dynamically re-arranged depending on the conversation [78]. The four areas of interest were application domain (e.g., *How did you come up with specific use cases or application areas?*), project management (e.g., *How did you run the project day-to-day? How does the project experience relate to previous experiences?*), requirements for IBM Watson (e.g., *Which preconditions did you have to fulfil to be able implement IBM Watson for this specific project?*), and impact on individual / human-computer interaction (e.g., *How do you ensure that IBM Watson or your application would be successfully adopted by business users?*).

All interviews lasted at least one hour, with persons involved in more than one project, proportionally longer. Seven interviews were conducted in English and 29 in German. All interviews were audio-recorded, transcribed (intelligent verbatim—the transcription represents recorded speech but without distracting fillers and repetitions), and offered to the subjects for review. To improve observer triangulation, we had two interviewers / coders supervised by three experienced researchers and two higher management persons from IBM. Observations were discussed in meetings throughout the data collection and analysis to support triangulation among the research team.

*Data analysis*. Data were coded in two rounds. The initial round was conducted bottom-up within four predefined areas of interest (*application domain*, *project management*, *requirements for IBM Watson*, and *impact on individual/human-computer interaction*) and yielded approximately 3000 relevant segments. There were two student assistants involved in data analysis; they employed iterative coding. This paper's authors controlled, corrected, and extended the coding during the process such that there is a common, coherent basis for the analysis. As all coders were bilingual, they coded the transcripts in the original language. The results of the initial round were summarized and discussed in two workshops involving the researchers and the representatives of IBM management in 2017 and 2018, as well as two workshops among researchers in 2018. The analysis of the initially coded segments, especially concerning project management, revealed that the interviewees engaged in frequent and intense sensemaking episodes. We identified a recurring theme of participants attempting to understand their experiences and emerging events. They frequently mentioned having learned new things that questioned their earlier assumptions and concepts. We note that the participants did not explicitly describe these processes as "sensemaking". Sensemaking emerged as our interpretation of the participants' statements concerning assumptions, questioning them, and establishing new meanings. Findings show what assumptions were made, what cues contradicted them, and what new meanings emerged.

The authors observed similar processes in their own AI-based project work with external partners. We noticed that the AI projects experienced significant tensions between expectations and reality, which required significant effort to resolve [79]. Our industry partners also lacked approaches to run AI-based development projects and exchange between industry players to develop insights for managing such projects. Inspired by those experiences, we revisited the previously collected data in a second round of coding to identify a reason for the tensions.

[6] www.ibm.com / watson; retrieved on November 9, 2022.

The second coding was conducted in 2019. It considered whole interviews again but followed a top-down process done entirely by the authors. First, we identified incidents of sensemaking in the interviews by attending to expressions signaling informants' establishing of new meanings. To identify relevant passages, we used references to *insights*, *learnings*, *expertise collection*, *experiences*, *unexpected events*, *breakdowns*, or *surprises*. We collated information about single incidents of sensemaking and treated them as the unit of analysis rather than attending to projects as a whole. The focus on sensemaking incidents fits the chosen theoretical lens. We found no significant differences between the projects, industry sectors, or AI use cases.

In the second step, we coded the information about each incident of sensemaking. The coding relied on the sequential view on sensemaking focusing on initial meanings, breakdowns or triggers for sensemaking, and recovery episodes. We also coded the sensemaking targets. The second-round coding yielded 350 coded segments used in this article. Exemplary coded segments are presented in Findings. After a gap due to the COVID-19 pandemics, the authors resumed work in 2022 compiling a first draft. The supplemental materials attached to this paper provide a list of codes used in the first round of coding, a timeline of the research endeavor and a list of strategies the authors used to assure the validity and reliability of the study.

## FINDINGS

Our findings are structured around the targets of sensemaking, a concept derived from our analysis. Interviewees discussed how their understanding of various targets evolved during the project. These targets were categorized into four topics: project, technology, data, and context. However, this division is for analytical ease; we will revisit how these targets are interconnected in Discussion.

Interestingly, we noticed no difference in the sensemaking needs of vendor and clients, although their initial understandings varied. For instance, both clients and IBM respondents reported significant evolution in their understanding of the technology, despite our assumption that vendors would have a better grasp. Similarly, while we anticipated clients' data would pose a greater challenge for vendors, many clients revealed that they began to understand their data through the projects.

We used the sequential and cyclic view of sensemaking to structure our narration of each sensemaking target and represent our coding scheme. Interviewees often described their evolving perceptions in a temporal order but sometimes the pieces belonging to a single sensemaking thread are scattered across an interview. We have organized their reflections according to sensemaking steps. We highlight the *initial expectations*, the conflicting cues that caused *breakdowns* in meaning, and the *provisional meanings* that emerged from the sensemaking cycle, which laid the foundation for recovery.

### Sensemaking of the project

The interviews revealed that despite the projects being considered as exploratory and learning opportunities, clients had to adjust their initial expectations throughout. We found two dominating expectations and several breakdown points. Ultimately, participants saw the projects as a chance for collaborative learning, involving intense, agile collaboration and potential failure (see Table 2).

Clients initially expected projects to resemble past collaborations, such as ERP system implementations. They anticipated *a clear division of expertise, responsibilities, and roles*: IBM would provide technological knowledge and developers, while the clients would provide context and subject-matter experts. Clear interfaces between partners were assumed. These expectations were based on *IBM's reputation, knowledge of software processes*, and *past experiences with external providers*. A project leader from a major Swiss bank shared her expectations:

> *I would say that IBM's market presence promises a great deal in terms of what they can do. (...) We need someone who knows the solution, i.e., the product itself, then we need someone who can guide us in the subject, then we need someone who knows references and knows best practice. (2C1)*

However, she came to realize that such expectations were inaccurate in the project. One problem was that *IBM was lacking trained personnel* because the technology was so novel. This lack caused the breakdown of her understanding about IBM and IBM workers' expertise, upending the client's notion of expertise:

> *We had a bit of a discussion about staffing, there were a few changes, and I could imagine that this is an indication that there are still uncertainties. Then, I had the impression that it is also difficult to get to the experts from Watson. A lot of them were based in America. And to get to them (...) was or is probably still quite difficult. (...) I don't want to accuse people of not knowing Watson, but I had the impression that they still had to do a lot of development in the background. (2C1)*

IBM's lack of skilled consultants was not the only issue. Clients found themselves with an unexpected division of labour, *taking over tasks which they expected to be done by developers or vendors*, like preparing data for machine learning tasks. For instance, a public transportation company representative was surprised to have to formulate questions for a chatbot solution. The shift in responsibilities was because the data needed to train a chatbot, question-answer pairs, are different from what other systems require and need domain expertise to generate. However, she found that this role positively influenced the project's scope definition. In this case, the lack of data and sensemaking thereof initiated sensemaking of the project and its scope.

> *[In the first test round] we only had 70 questions. (...) We simply had to generate a lot of additional questions with our internal resources. (…) At the beginning we had the hope that there would be something more coming from IBM, but—as it turned out—it was actually our work to generate those questions. Accordingly, the scope of the project sharpened a bit: we started reflecting "can we ask this question at all?", (...) the developer was just there, and he said to us "ahhh, this is more difficult now", "too expensive", "is not worth implementing", "too much effort for very little intent[7]", and so we automatically came up with the scope. (5C1)*

Yet, IBM representatives also admitted facing challenges due to the project's unique environment. They cited *the project's nature* (e.g., innovation vs. outsourcing vs. internal development project), *fluctuating cost structure* (e.g., increased time and effort needed to iteratively retrain models if goals were not met initially), *and internal expert shortage* as issues. An IBM project manager, working with the mentioned public transportation company, concluded

[7] In natural language processing for a chatbot, 'intent' refers to an intention expressed by the speaker as recognized by the ML engine.

that a dynamic and collaborative approach is essential to manage inherent uncertainties.

*A very agile cooperation is needed, really, an extremely flexible cooperation, if you want to get the whole thing through. (...) Because there is just so much learning and training involved. And if you charged the client the whole project management and all time spent, then the costs would simply increase massively. (...) So we could not even calculate everything because the project budget would simply explode. (...) And there was yet another challenge I did not expect: to have the right skills and right people on board from IBM at your disposal. Because those specialists are apparently in high demand as they are still a bit rare on the market or in our company. (...) So it's about learning and flexibility. (5V1)*

The projects were considered more dynamic than people were expecting or had experienced. An IBM professional with four years of experience depicts it as follows:

*There were always changes to the planned. People had to react really quickly. Every Friday after the meeting in the morning, we said in the afternoon: 'OK, we'll do this differently, we'll do this, we'll do that, we'll leave that, what else is needed?' That was agile, there were changes every day. (...) That was the most intensive project I have ever done. (14V1)*

The previously mentioned project leader from the Swiss bank puts it all together. She indicated that she originally expected that the project will be about getting a product that can be put to work. However, based on the experience that *goals and ways to achieve them are unclear*, she arrived at a new perspective, that the *project was about learning,* and *failures are acceptable if they produce useful learnings*:

*In a normal IT project, if you want to upgrade the version, it is clear, we go from version 1 to 2, the target is clear. (...) Here [in Watson projects] a great deal of effort is required to find things out, which means you don't just buy the product from the supplier, but rather a whole bouquet of services with a great deal of uncertainty. (...) We first thought that the solution would be more established. (...) But then we had to realize that many questions are still unclear, even for the IBM people. Because it's the first time that they do something in exactly that way. And that's why it was a learning process. We really had to realize that we are here in an environment where everything is shaky. (...) It's good too, that was an important experience, but that was new. (...) [I learned that] it's good to push the story to the end, even if you realize along the way that it won't be as successful as you thought. So it was still worth the effort, even if, as I said, not much came out of it product-wise, but there were still learnings. So, 'plod through!', I'd say. (2C1)*

Some participants suggested a strategic approach for resource allocation and scope due to the project's uncertain nature. They recommended using trial-and-error in acceptable failure zones to identify optimal solutions quickly, improving project risk management. A client-side project leader from a major B2B insurance provider shared learnings from a service desk support project:

*From a project management point of view, it's a nightmare to invest money in things that are then discarded. I think one of the lessons learned is that we bit off a little too much. Smaller bites would be better from a project management point of view, also from a risk point of view. (...) We have already tried this [dividing in 'smaller bites'] and I think you can do it even better if you pay more attention to the risk, so under the motto 'fail fast'. (...) Reaching the point even faster with even less effort where you say, 'this won't work, I'll throw it away'. That can be done even better. We waited too long for 1-2 things and spent too much time to say, 'This is useless.' (17C1)*

TABLE 2
DEVELOPMENT OF MEANING CONCERNING AIDP-BASED PROJECTS.

| Initial meanings | Sources of breakdown | New provisional meanings |
|---|---|---|
| - Perception of IBM's workers as experts based on IBM's reputation.<br>- Ideas about distribution of expertise, responsibilities, and roles between vendor and client in outsourcing projects based on experience of client-vendor projects (e.g., ERP implementation) or knowledge of systems development. | - IBM's lacking accessible trained personnel.<br>- Necessity to take over tasks associated with a different role in outsourcing or collaboration in IT projects.<br>- Changes in project goals, project structure, or project timeline.<br>- Missing clarity about the goals and ways to achieve them. | - The projects are primarily about learning and not only about delivering a product. All members in the project are learning.<br>- The projects need to be highly dynamic and collaborative to accommodate for uncertainties.<br>- Project scope needs to grow in small steps.<br>- Taking over some tasks provides possibility for collective specification of the scope.<br>- Failing is acceptable as long as there are learning effects. |

## Sensemaking of the technology

Clients were expected to be engaging in sensemaking around the technology. Initially, they anticipated the technology to *be an upgrade to systems they knew* or *like what was presented in IBM's sales events and media coverage*. However, upon starting the project, they found that Watson's capabilities did not match their expectations as *Watson could not handle the desired use cases*. An insurance company representative initially saw Watson as a superior alternative to their existing IT, replacing all of its functionality. However, she soon realized that conventional tools were still needed, e.g., for handling data, with Watson providing only incremental functionality.

*The expectation really was that you can take the next step with these cognitive approaches and really make it more efficient, generate better insights, and simply manage the sheer volume of data. (...) Watson turns out not to be an application or a process or the like, but it really is a portfolio of different products. I think what Watson does for people, is basically stupid work, really reading through documents and maybe finding the relevant passages or having an view of everything. (...) That means we had to build some support around it, for example, to make a connection with our document management system. It was actually one of the selling propositions of the tool that it would be able to deal with the documents out-of-the-box. But that never worked properly. That means we basically built a little something to make it work. I have to say, it was a bit disappointing to us. One could have expected better. (8C1)*

Asked about whether there were essential differences concerning the expectations compared to legacy projects, she offered the following explanation indicating what IBM could do to help clients like her:

*The customer's expectations are perhaps higher [than in other projects]. IBM has to be much more active in managing expectations, because the client says, 'yes, I've seen Jeopardy on TV, but you're taking forever here'. That means IBM really has to explain to the client what cognitive computing is and that it doesn't just work like that, but that you also have to do a lot of this training step and feed the system with knowledge. (8C1)*

Overall, this case indicates that clients frequently started with expectations based on marketing established around Watson's capabilities and expected that those capabilities can be transferred to their work context with little additional effort. However, even working solutions that IBM developed in earlier projects could often not be easily transferred between companies, e.g., due to data and domain dependency. When Watson was not able to handle specific use cases from client's specific business context, they had to accept that integrating Watson in their own technical or data context is necessary for it to work at all.

Another source of breakdowns in sense were *frequent changes of the platform and its components*. This aspect was equally surprising to the clients and to some IBM employees. An IBM-side project manager stated:

*That [in Watson projects] is where the relationship of trust is essential. And this was also burdened by the fact that the entire field was very dynamic (...) Very often different statements were made by our side about which products were suitable. (...) The product sets themselves were very dynamic and continue to be highly dynamic. (…) There were always different statements about the availability of the*

*languages, for example. That of course did not improve the relationship of trust, because we, as representatives of our brand, were a bit insecure and could not do so much. (…) One must compensate for it through particularly stringent and clear communication, and one must manage this communication. (2V1)*

The structure and dynamic of the platform were indeed causing confusion on the clients' side:

*For us Watson was originally simply a product, but then we had to learn there are different components, Explorer, Advisor, there are ticketing tools somewhere and so on and everything together would then give a product. And then we lost the overview a bit on the product side. (2C1)*

Despite their initial expectation of Watson to be a ready-made product, clients and IBM employees had to realize that it is *a very dynamic development platform with uncertain structure and scope*. IBM representatives openly acknowledged that this created issues and the need for extra clarification, amplifying the uncertainty stemming from the probabilistic and complex nature of ML-based reasoning.

A unique feature of current AI tools, as in Watson, is their data-driven, probabilistic approach, which can result in errors and uncertainties. Despite being aware of this, clients were still overwhelmed by the uncertainty level:

*Normally [a product] means for me a standard solution with known technology and a known environment. Here, the starting position was actually quite different, because you didn't know what you were actually talking about. (...) But if you really want to integrate it, there are completely different challenges, in contrast to a standardized tool, where you can really show the input mask, 'these are the metrics, these are the fields, this is the report', you can show everything, and with such a new technology and with a cognitive story, what comes out at the front must be determined at the back, and therefore the uncertainties are actually much greater, the unknowns were greater than I expected. (2C1)*

Respondents found that some errors in the output from Watson tools could not be rectified by code alterations but instead needed costly data processing and retraining, with no guarantee of success. They found that *extra effort doesn't always enhance performance*. The head of a product unit at a prominent Swiss telecom company, who was the client-side project lead, elaborated on this difference. The project's aim was to enhance customer-support-center processes by automating incoming email triage and aiding response efforts. When asked what surprised him most, considering his past experiences, he responded:

*I think the main difference is with regular IT projects, basically, you have a process and I think it's more a question of time and money whether this is going to be implemented the way you specified it. With AI and cognitive project, the biggest difference is whether the program or the model will find patterns to allow you to, basically, cluster the data you had in meaningful data sets. Do these patterns exist, so that you can really classify your data in a way which is useable afterward? And that you don't know until you make this exercise. So, I think that's the real difference. You can start the project, but you are not sure that the results will be something useable at the end. (1C1)*

TABLE 3
DEVELOPMENT OF MEANING CONCERNING WATSON AS TECHNOLOGY.

| Initial meanings | Sources of breakdown | New provisional meanings |
|---|---|---|
| - Perception of extraordinary Watson's abilities based on sales events or media coverage.<br>- Knowledge of AI and ML techniques.<br>- Watson as an upgrade or end-to-end product, similar to other software systems. | - Limited knowledge of IBM's personnel about Watson, its roadmap, and its abilities.<br>- Frequent updates to the Watson AIDP (new modules, names, structures).<br>- Watson's inability to handle particular use cases. | - Watson is a flexible, emerging AIDP with uncertain structure and scope.<br>- More communication is needed to accommodate for the dynamic nature of the platform.<br>- Additional effort does not directly relate to the system's performance.<br>- Integrating Watson in the technical or data context is necessary for it to work at all. |

Overall, the client and IBM employees both engaged in the sensemaking of the technology. Whereas the clients' representatives seemed to expect something novel, they are frequently overwhelmed by the differences between what marketing and media coverage of Watson led them to believe was possible and what was indeed possible with a given tool in the chosen use case. Both sides responded by intensifying communication and collaboration to compensate for the uncertainty created by the "continuously becoming" technology, as summarized in Table 3.

## Sensemaking of the data

Third, the analysis showed a strong need to understand the data to make sense of Watson's behavior. Training models is crucial in Watson projects, making data a key project asset. Dependency on the availability and quality of data was the key characteristics of projects as identified by participants. IBM representatives were prepared for potential surprises regarding the data, highlighting the need to comprehend the data to understand the client's needs and capabilities. An IBM interviewee summarized it as follows:

*Normally we have an opinion and say: "Let's do the little things first to get a feel for the data" (17V1).*

Clients confessed to only focusing on their data after the project commenced. While they were aware that data was necessary and believed it was available, they often *overestimated its value and size for machine learning components*. The mentioned employee of a Swiss telecom company explains that the *amount of available data* needed to be increased:

*So, we had a large data set (...) I think we had something like 10 thousand and 50 thousand emails we were providing (...) At the start, I think, it was 10 thousand but then we managed to get more emails than what we originally thought about. So, basically, we had the set of emails for them to train. (1C1)*

Looking back, the client admitted that exploring data earlier would have improved the project. The problem was not just the size, but also data *format, which Watson could not process,* despite *manual review* suggesting otherwise:

*Well, the data collection part was much harder than what we thought for several reasons. Because the data was not ready to be addressed with such a technology. Like a simple problem was all the emails and answers to them were saved in a database, but there was no key relating a question to an answer because there was never a need to try to link the two. [As a human] you could learn from it anyways. But then we had to develop some heuristics to recreate that link afterward. (1C1)*

The nature of the data also created issues, for instance a skewed data distribution, discovered mid-project, created problems for model training, though not for routine use. Further, the interviewee found the *data's inaccessibility* problematic, causing project delays. He reiterated the importance of early, thorough data pre-processing.

*Additionally, these databases didn't have an interface to retrieve a lot of data in one go, because there was no need for that before. (...) So, in retrospect, I would have spent more time earlier to get the data, familiarize ourselves first with the data, because sometimes we had to discover the data at the same time as the vendor. And that was not so comfortable I would say. (...) I think, the big majority of emails we have are questions about invoicing. The distribution is probably something like this. So, it would have been also interesting to see. Perhaps, I would have taken cases where we have lots of emails, cases where we don't have so many emails, and see whether the performances are different or not. That would have been another dimension we could have introduced in the project. (1C1)*

Besides issues with data handling and access, breakdowns were triggered by the system's output based on client-provided data. This was evident in a project between IBM and a large pharmaceutical firm. The project aimed to expedite identifying potential partners, such as startups or hardware providers, crucial in the R&D-dependent pharmaceutical industry. The IBM project leader shared how presenting Watson's outputs to the client highlighted the importance of data for their solution:

> *They did the initial mapping of the companies, for all the 3000. This is where we could have brought some more capabilities in. Their data was 3000 rows of spreadsheet. This was also not as complete as we expected. They had a lot of issues in their data. For instance, there were some companies that kept popping up and they were like: "Why is it showing up?! That doesn't make any sense. Your application doesn't work". Then we all went back to the source data and that output was sitting directly in their spreadsheet. They were like: "Ah... Okay." There were some quality issues on their side. (20V1)*

Interviewees determined that data issues were not specific to a use case but *tied to an organization's culture or an entire industry*. In a project where IBM and a large Swiss bank aimed to automate contract analysis for risk identification, the subject matter expert shared insights on project progress, data quality, and data management practices.

> *So basically, we learned it as a three-step curve. It basically went in three waves. The first part is using cognitive intelligence to improve your data. The second component is then using cognitive intelligence in the data itself to improve it. The third part is true cognitive computing. I think that's the value chain. (...) There's a lack of understanding within the banking world, or within [The Bank] especially. Everyone thinks you can achieve the gold at the end of the chain. Whereas your data is not capable of doing that. (...) With Watson, we needed a certain level of structured data plus IBM's instruction. Then you build your algorithms on top of that. [The Bank] is not at the point where we have completely good data quality and I think most banks will say the same. (3C1)*

Overall, vendor employees were often ready for data challenges, while clients were regularly taken aback by their company's data quality. Many assumed they had ample relevant, accessible data. However, using this data with Watson revealed issues: either the data were unusable, or the output was incorrect. Recognizing these shortcomings, clients started to address data quality both reactively and systematically. Table 4 outlines this process.

### Sensemaking of the system deployment context

Project members considered the wider context of system deployment. They ensured the solution's relevance to its specific use context, but the broader context became significant later. Initially, teams aimed to enhance the work of certain client employees or customers. However, these stakeholders often had mixed or negative reactions to the system. This led to the recognition that *greater stakeholder integration in the development process is necessary,* not just during conception but throughout. Also, it is necessary to *address their AI-related fears*.

Interviews suggest clients believe their current processes are sufficient to use Watson, and their employees will accept it. However, as noted by IBM employees, there might be interest from the to-be-supported personnel might be lower than expected. A project with a large pharmaceutical company highlighted this mismatch between client expectations and user reactions, as described by an IBM representative:

> *I put together 70 people on their list of pilot users, which was a bit extreme. They wanted to make sure they hit many different areas within the [Company] space. (...) Then people were asked to really start to use the tool. The reality was again somewhere between 10 and 20 people used the tool. Of which probably only 5 really used the tool. Many people that had access never logged in. (...) These people were supposed to really use the functionality to see if this does help their job. It was a very difficult area to put an ROI to, return on investment. (20V1)*

IBM team members highlight the need for an extensive change management strategy due to Watson's potential to alter key work aspects and introduce new tasks. Further, AI has the capability to support or automate not just routine tasks but also complex tasks that require intellectual capabilities, such as the assessment of the risk associated with a contract. This expanded scope of application might create new expectations for the users and disappoint them more if those expectations cannot be met. In a project, a Watson-based solution was developed to aid service desk employees. The IBM project leader discusses the risks of inadequate change management:

> *This brings us back to the topic of communication and change management. You can't just take an old process and put cognitive computing under it, that doesn't work. (...) People just don't go and say, 'Yes, now I'm going to rebuild my process,' based on an application I don't know, a tool I don't trust anyway, and a technology that is primarily marketing.' (...) We now have an Assistant View where one can go to Watson and then go back to the old process. If you do it that way, you don't use the full power of Watson. (...) If then the first 4-5 times it's not so good, then you just go back to your old, main branch of the process and think 'This is too stupid for me'. It takes a lot of change management to somehow force people to take the sub-branch of the process with gamification or whatever. (17V1)*

This IBM member concludes that a big-bang launch could lead to effective software and processes. However, she also acknowledges the risks associated with her rollout strategy assumption:

> *If you can manage, "bang" old process out, here's your new interface that we designed with you and you have to go through this Watson thing, then you'll have a solid function within 3-4 months. But this was theory. The current practice is very much rooted in the human nature, now I get it. (17V1)*

Aside from low adoption, another trigger to question initial assumptions was the *client's staff fearing the new technology*. This is particularly striking for AI, as people have strong preconceptions induced by popular media or culture. For example, a representative from a Swiss manufacturer, responsible for internal stakeholder relations, shared her experience with a project for a research department to identify potential industry partners. This project resulted in a more precise search engine than previously used. However, the stakeholders reacted emotionally due to information from culture and popular media:

> *People are always scared and think of 'I, Robot', 'Terminator', or whatever. These films suggest that machines are taking over the world, that can happen—I'm not sure either. But at the moment it's different. Instead of taking the machine as an adviser and we [humans] making the final decision, it gives them fear that the machine will make all the decisions. And that's not true! (…), I can see it in people's eyes and body language, and then we have to take a step back. We're still too early in the game for that. (...) I think that's a trend right now: everyone does that, so they do that too. (14C1)*

The same interviewee, however, indicates that providing tangible prototypes helped with those fear:

TABLE 4

DEVELOPMENT OF MEANING CONCERNING TRAINING AND TESTING DATA.

| Initial meanings | Sources of breakdown | New provisional meanings |
|---|---|---|
| - Notion of possessing much data or awareness of data collection efforts.<br>- Manual review of available data. | - Watson cannot deal with the data as provided.<br>- Application outputs are not comprehensible or assessed as clearly wrong.<br>- Data is not easily accessible. | - Preprocessing, structuring, and re-integrating the data is a prerequisite and a necessary part of the project.<br>- Output needs to be checked against the data.<br>- Data quality is a matter of (organizational) culture. |

> *The learning is that visualization is very important, so that you can show people a thing in a short time. Otherwise, they don't believe you, because it's all very, very abstract for them. (14C1)*

The interviewees reported that the fears were often abstract and not grounded in reality but still problematic. They adjusted their approach to address these fears, *making things more tangible* and *implementing a gradual rollout process*. The Swiss division of a large material science firm aimed to create an app to serve as a central hub for searching and interpreting company data across various platforms. The project leader from IBM discussed the rollout strategy used:

> *I don't think users see what we're introducing now as a threat, but rather as a useful addition or support for their work. (...) it helps us now, or it suits us, that we were not taking such a big-bang implementation approach, but rather approaching it relatively slowly. The users slowly get used to it and gain confidence in the application, solution, or technology. We encountered more skepticism at the beginning when we were still discussing it at a high level. Now we are in the implementation. Now it's tangible, they can see it, they can approach it and they can see that it's not threatening at the moment. (13V1)*

The clients and the IBM expected that they will need to make sense of the context of use. However, the societal context's impact was unexpected for clients. Fears often surfaced through references to popular culture and were amplified by media coverage. IBM representatives were more cognizant of these concerns and aimed to address them. To understand these fears, the team sought to engage more frequently with users and stakeholders who might be affected by the solution. The process is outlined in Table 5.

### General patterns across sensemaking targets

Table 6 outlines shared aspects of sensemaking targets, despite their distinct characteristics. Project participants drew from various sources, leading to inconsistent expectations. Some referred to standards, others to prior project experiences, which varied across and within organizations. Understandings of AI's nature and potential also differed. Initial expectations were often based on public data or personal experiences. The absence of a shared reference point could cause misunderstandings and necessitate meaning negotiation among individuals.

Breakdowns are triggered in various ways. Some are related to the desired product, with partners questioning expectations if the product does not meet them. Actions of team members might contribute to breakdowns if their behavior is inconsistent with the perceptions of their abilities. These breakdowns may even lead to questioning the capabilities of one's own company, e.g., when it turns out that it lacks the necessary resources. Finally, Watson can disturb existing meanings too.

Interviewees acknowledge the complexity of the projects, with numerous known and unknown factors needing attention. They find the effort-outcome relationship unclear, complicating resource management. The project's opaque nature can confuse decision makers, making resource acquisition challenging. Generally, the interviews emphasized the projects' explorative and learning nature.

TABLE 5
DEVELOPMENT OF MEANING CONCERNING THE CONTEXT OF USE.

| Initial meanings | Sources of breakdown | New provisional meanings |
|---|---|---|
| - Employees will accept Watson and are willing to include it in their work processes.<br>- Existing work processes and employee training are sufficient; system should support the process. | - Little interest from employees who should be supported with the intended application in the future.<br>- Fears fueled by public discourse or individual situation without direct relevance to the particular solution or project. | - Making things tangible helps communicate with stakeholders and reduce their fears.<br>- Slow, step-by-step rollout as a way to accommodate for concerns of the stakeholders. |

TABLE 6
SUMMARY: DEVELOPMENT OF MEANINGS IN AIDP PROJECTS.

| Sources of initial meanings | Sources of breakdown | New provisional meanings |
|---|---|---|
| *Public:*<br>- Depictions of Watson or IBM from media and sales events.<br>- General depictions of AI in media and public discourse.<br>- Examples of earlier applications.<br><br>*Individual / Personal:*<br>- One's perception of the own organization.<br>- One's perception of existing work processes or practices.<br>- Perception of the partner organization.<br>- Own experience from previous projects. | *Product-related:*<br>- Low, inconsistent, or random performance.<br>- Incomprehensible, unexplainable output.<br>- Non-transparent relation between invested resources and outcome.<br><br>*Team-related:*<br>- IBM's lack of adequate human resources.<br>- Obscure actions of IBM concerning Watson.<br>- Obscure actions of clients concerning data.<br><br>*Platform-related:*<br>- Frequent changes to the platform and lack of knowledge about those upgrades. | - Learning as the main activity: based on the application of the preprocessing, training, and testing machines learn to complete a task and humans learn about the data, the machine, and the task.<br>- Performance depends on the quality of data and fit between data and algorithm.<br>- Performance is not directly related to developers' skills or time spent developing.<br>- Making things tangible could help but is extremely difficult because reasoning is an abstract process and improvements in reasoning hard to represent.<br>- AI development involves unknown unknowns leading to a need for spontaneous reconfigurations of the projects. |

## Discussion

The object of this study was to describe the process and targets of sensemaking in AIDP-based projects, using evidence from IBM's efforts with their Watson AIDP. The analysis of respondents' experiences with Watson projects shows that the AI components increased the complexity of the projects, which exceeded project members' expectations. This complexity caused frequent breakdowns of meaning that left the participants temporarily without frames they could act on or use to assess the project. When new frames were established, new meanings propagate through all areas of the project. This process requires time and resources, causing delays and increases in costs. It also makes the client companies conclude that much more learning is necessary before they can purposefully or productively participate in AIDP-based projects. Our findings have implications in three domains: the areas of sensemaking, the process of sensemaking, and the implications of project complexity on sensemaking.

### Targets of sensemaking in AIDP-based projects

We identified four targets of sensemaking in AIDP-based projects: data, technology, context of use, and the project itself. All four contribute to the complexity of the endeavor and become subject to reflection throughout projects.

AIDP-based projects emerged to be about making sense of *data* and about what technology, as opposed to a human, can derive from this data. Whereas systems projects have always involved decisions about data storage, management, or access, AIDP-based projects bring a new perspective to this topic: are the data sufficient and adequate for training, to what extent are they structured or dynamic, what are the quality attributes of the data? Providing adequate data is the primary possibility to improve the performance of the desired application. This concern moves the initial focus of development away from choosing the right algorithm or creating bug-free code to data selection, structuring, and generation. The need for sensemaking around data results to a large degree from *inherent characteristics of current AI technologies*, many of which rely on (large) amounts of data to learn from. Further, sensemaking of data is only possible in combination with AI capabilities. Only through combination with an algorithm or a model, users can learn something about their data that *makes sense* within the context of AIDP-based development.

We also observed that sensemaking around *technology* in AIDP-based projects follows its own path. For instance, project members found that it was not possible to assume the transferability of Watson's abilities from one application (e.g., question answering in a general-knowledge quiz show) to another application (e.g., question answering related to a particular railway station) even if the tasks might seem very similar on the surface. Instead, sensemaking about capabilities needs to occur anew with each change of the data set, the context, or the task.

The technology provided as an AIDP generates an enhanced necessity for continuous sensemaking. The Watson AIDP was undergoing steady development such that learnings one made even only a few weeks ago might no longer be applicable. This rapid pace of change—which seems to characterize the current state of AI technology more generally—affects clients, developers, and IBM consultants who need to adapt to the changes on the go. A project might need to develop a chatbot from scratch, while couple of months later the AIDP might be extended to cover this capability out of the box. This possibility leaves clients concerned about the necessity of some investments.

The need for sensemaking around the technology also results from the *hype* that surrounds IBM Watson as well as the *inherent characteristics of AI* as a paradigm. The participants entered projects with assumptions about AI and, specifically, Watson based on how it was portrayed in the news or in IBM's presentations. Those beliefs were then invalidated by, e.g., lack of capabilities attributed to Watson. Hype was amplifying the initial expectations even more creating the expectation that Watson might essentially transform the organizations, which was invalidated during the specification of use cases.

Because of the difference in the role of data and the nature of the technology, the traditional notions of what SE *projects* are about did not seem to apply entirely to AIDP-based development. Sensemaking about the project and the partners intensified. While one might tend to assess a partner based on the performance of their output, applying this strategy to AI-based projects can be misleading. The client might be tempted to attribute bad performance to the application and the developer team, though the provided data might be the reason. Accordingly, a clear-cut responsibility structure and performance-based evaluation is problematic as a guide for the partners' interaction. In the observed projects, the participants settled on collective learning as a frame for collaboration, but this view might not be applicable in all commercial settings. The sensemaking about projects was primarily driven by the *novelty* of the technology and, thus, the novelty of the project type. IBM and clients both lacked sufficient experience and adequate expertise because there were not enough people who had to deal with this technology before. People instead resorted to other frames to inform their sensemaking [18].

Ultimately, sensemaking is crucial in comprehending a client's *context of use*, particularly for providers [61]. This process becomes even more vital with AI technologies, given increased media coverage and metaphors like 'robot' or 'terminator', requiring project members to understand them and strategize accordingly for AI deployment. Further, AI-based systems can be applied to jobs that previously have been less affected by automation. The probabilistic nature of AI requires substantial shifts in employees' work approach compared to business process automation. Hence, sensemaking encompasses not only analyzing employees' work, potential changes in their practice, and implementation strategies, but also envisioning the future of work more generally. The need for sensemaking of the context was complicated by the *hype* around AI in general. Since virtually everyone has heard about it before, affected individuals would frequently have made up their mind before even interacting with AI. Revising those beliefs needed additional sensemaking from participants.

| Expected need for sense-making | | IBM | |
|---|---|---|---|
| | | Yes | No |
| Client | Yes | Context | Technology |
| | No | Data | Project |

Fig. 1. Expectations on the need for sensemaking in different areas.

Interestingly, the interviews reveal *waves of sensemaking* initiated by cues starting from data or technology but then spreading through the other areas of sensemaking. For instance, in one of the projects, the data initially provided included 70 questions for a chatbot to handle but these were revealed as insufficient in the first round of testing with Watson. The client's initial understanding of data and technology broke down. They recognized that more data would be necessary and, because of the need for subject matter expertise, that new data would need to come from the client. However, this new meaning was incompatible with the client's understanding of the project and the distribution of responsibilities: they initially did not consider manual work with data their responsibility. The understanding about project roles broke down and the client started enacting a new, more engaged role in an overall learning project. However, when IBM did not involve them in training the model—as would be appropriate given the newly-established learning character of the project—this meaning broke again. This example shows how a breakdown of meaning in one area, technology (incapable of learning from too few data points), propagates to a breakdown of the meaning attached to data (existing data insufficient, new data must be generated), which in turn propagates to other targets, in this case, the project (client responsible for data generation). The interviews point to multiple cases of such interdependencies, which suggest that sensemaking in each of the four areas is tightly related to others. The sensemaking of a complex project is itself a complex phenomenon and requires a careful approach.

The study uncovers socio-cognitive processes involved in building up and overcoming the complexity of AI-based applications. Specifically, it shows that uncertainty and breakdown in one area might transfer to other areas, and that the areas are tightly interconnected. This connection implies that the challenges outlined in previous literature [4], [9] need to be addressed as a whole rather than one-by-one. For instance, extending the project team to include a new category of specialist, a data consultant [18] to bridge the business context and data, will have positive and negative implications for other areas that need to be assessed. A data consultant might care for increasing the fit between the available data and business needs through selection and pre-processing, but this processing might simultaneously make the data even more opaque to the remaining project members. This observation adds to SE literature on sensemaking as a necessary step to interpret the application context of the software product [80], [81], or as a prerequisite for mutual understanding in dispersed or inter-organizational teams [61], [82], [83]. This study advances this research by applying sensemaking to the context of AI application development.

### Process of sensemaking in AIDP-based projects

Turning to the processes of sensemaking, we found similarities across the identified areas concerning the source of the initial meanings and how the declared character of the projects (i.e., innovation, exploration, research) impacted the process of sensemaking. The client's initial understanding of the abilities of Watson, capabilities of IBM, or the notion of unstructured data frequently referred to IBM's marketing and sales materials. The IBM-side informants reported that they were often overwhelmed by the expectations generated by this positioning and tried to set more realistic expectations upon starting the project. However, the clients only fully realized the limitations when confronted with sometimes unsatisfactory outputs, the amount of manual work involved in producing those outputs or the realization that the Watson offering was still emerging. Participants on both sides faced innovative technology, novel perspectives on data, new project structures, and new concerns from the future users. Thus, when their initial meanings failed, they embraced the explorative character of the projects and became more open to iteratively establishing new meanings.

However, this shift in perspective was not equally easy for each of the sensemaking targets. Clients were prepared for the novelty of technology but frequently did not expect that their understanding of their own data would be incomplete. Similarly, they were prepared and looked for relevant cues concerning the vision and the context of the application. However, their understanding of how the project should be run was initially settled. In other words, they were prepared for sensemaking about the technology or the context, but the need for intense sensemaking concerning the project and the data surprised them. This surprise is reflected in their reactions, such as the initial reluctance to accept that there might be mistakes in the data or the concern about taking a role that was incompatible with their predefined identity. Similarly, the IBM representatives frequently expressed that they were unprepared for difficulties concerning the project management consequences of the availability of personnel or financial issues and the restructurings of the AIDP technology itself. Yet, they knew they would need to make sense of the context and the data. Figure 1 summarizes those expectations.

To interpret this finding, we draw on the concept of mindful organizing by Weick and Sutcliffe [13]. Our analysis suggests that mindfulness was not a general attitude but differed in terms of focus: clients were mindful concerning technology and context, whereas the vendor was mindful concerning the data and context, and neither was mindful about project organization. The results point towards a phenomenon of *selective mindfulness*: Project members were prepared for intense sensemaking for some targets and thus more flexible and more receptive to novel meanings in those areas. However, for other targets, they relied on predefined meanings. Only when it was inescapable to change the course of action because of the incoming cues did they establish and enact the new meanings.

This finding adds to the notion of mindfulness for SE projects [13], [84]. Individuals and organizations frequently focus on what is explicitly new, like technology or application context, and are mindful about those aspects, while staying mindless about what they think is known. We propose instead a notion of *balanced mindfulness*: In SE involving novel technologies, project members need to attend to all aspects mindfully. They should look out for cues from all sides of the projects and incorporate them early rather than relying on predefined meanings. Taking any aspect for granted or indisputable could negatively impact a mindful reaction in other areas. Being aware that there are multiple targets for sensemaking is the first step.

A second observation is that sensemaking in AIDP-based projects involves multiple actors beyond just the client and provider. The current study indicates that sensemaking in AI-based projects is a multilateral and collective phenomenon: the cues essential for sensemaking occur not only on the line between IBM and the clients but are produced by various stakeholders on both sides, including marketing and sales department, future users of the developed application, or AIDP contributors who are not part of the project. Even the technology or data themselves produce important cues, thus entering the communication between the actors. Furthermore, as signaled by the interviewees, sensemaking about one's own organization and its contributions is equally crucial and influences an individual's actions. This outcome calls for a collective perspective on sensemaking in SE projects [84]. Sensemaking is a multilateral phenomenon with various individual stakeholders taking various roles from all organizations.

Project participants also cannot rely on cues coming from a partner being consistent. We observed attempts to establish a distance from cues produced by other members of the same company, like the IBM representatives who contradict messages produced by marketing and sales or who openly criticize how Watson AIDP was managed. Those instances of collective mindfulness observed in the data ignore organizational boundaries and loyalty. We conclude that an analysis along the lines of provider-client or developer-client configuration [37], [61] might depict the interaction too simply. We call for more attention to the multilateral characteristics of collective sensemaking in SE.

In summary, the current study provides a differentiated picture of sensemaking in provider-client relationships. First, it shows how a mindful approach helps the participants accept new meanings more easily. Second, it suggests that mindfulness is not a general attitude but is rather target-dependent. Third, it indicates that sensemaking in inter-organizational projects goes beyond the client-vendor divide and happens between various actors within and across the organizations. Finally, sensemaking crosses various dimensions of the past and the future.

### The complexity of AI-based development raises new problems for sensemaking

Our study addresses the complexity of AIDP-based development and its impact on sensemaking. AIDP-based projects face common challenges such as vague requirements or budget issues [67]. However, they also introduce new complexities that demand and complicate sensemaking. The dependencies between various elements of the project become increasingly difficult to predict. As AI can yield unpredictable and complex results, traditional project methodologies and meanings established in those projects are often inadequate. We indicate that project members frequently started with assumptions modeled after the conventional software despite their awareness that AI is different. During the project, however, they had to update their believes about the implications AI has on SE.

Our study confirms the unique challenges of AI-based development previously noted [9], particularly in software testing and quality. The interdependency of system components complicates accountability for poor performance. Often, clients felt IBM overpromised, while at the same time, limitations of their data contributed to subpar results, both affecting mutual trust. Due to the lack of test data and high costs of systematic tests [35], [36], *ad hoc* assessments were frequently used, shaping perceptions about the technology and project. AI-related challenges in software testing and quality lead to mutual blame for project failure, potentially manifesting as criticism of the AIDP or provider, or highlighting the client's lack of preparedness.

Our interviews confirmed the challenges of Requirements Engineering in AI, often due to employees' misconceptions about AI, influenced by media and public discourse. IBM's sales events, demonstrating high-performing Watson applications, often led clients to form unrealistic expectations, similar to what was reported in earlier studies [9], [40]. IBM project leaders attempted to moderate these expectations, causing potential confusion between management and project-level expectations. This mismatch often resulted in requirements being discarded in favor of a learning approach. This outcome highlights the complexity of requirements management in AI projects [41], [42], necessitating consideration of public AI discourse and the need for ongoing reassessment of AI capabilities in the projects.

The analysis reveals that the technical complexity of AI-based solutions presents challenges in Software Design, Construction, and Maintenance [44], [45], [53], which—in turn—necessitate increased sensemaking. Rapid changes in Watson as an AIDP also caused confusion. The lack of knowledge about the system's capabilities among IBM members negatively impacted client perceptions. However, this issue is not unique to IBM. As AI models and tools offered by major providers like IBM, Google, and Microsoft continue to evolve, developers must stay updated with these changes. The growing ecosystem of tools and APIs may overwhelm developers, who must constantly learn new capabilities. The dynamic nature of AI, the inability to predict a tool's performance, and the CACE principle will transform the developer's role and client expectations. A developer's willingness to learn may be more valuable than thorough knowledge of an AIDP. We conclude that AI's inherent nature and its delivery via AIDP require constant sensemaking and reassessment of developers' skills.

Finally, the analysis highlights new challenges related to sensemaking in SE Management, Configuration Management, and Professional Practice. Developers often face a complex ecosystem of models and components that require continuous sensemaking, similar to understanding the usage context [16], [28]. Our study confirms this, emphasizing that AI system development demands more intense and frequent sensemaking than conventional systems. This sensemaking extends beyond users' context, model ecosystems, or AI development environments [16]. It includes project and data sensemaking and understanding AI's inherent nature. This process occurs throughout the project, triggered by various difficulties or observations. From the perspective of the interviewed project leaders and experts, sensemaking is not just necessary but also a significant source of progress, generating learnings and informing future steps. This result has implications for managers who need to consider sensemaking activities in project planning and execution, and who engage in the process themselves to adjust their own understanding of success and progress measures. Appropriate SE Models, Methods, and Processes can aid management in this regard. Consequently, we urge the SE community to engage in providing adequate support for project leaders.

### Towards sensemaking as part of the AI development process

The inherent complexity of modern AI [7], [8] demands sensemaking from all AIDP project members. This complexity impacts the deployment of AI capabilities. Due to the recent surge in AI popularity and lack of specific paradigms for development, deployment, and maintenance of AI systems, we reference MLOps [46], [47], [48], [50] and pipeline architectures [50], [51]. While these models recognize the iterative nature of improvement and need for system optimization over time, they overlook the requirement for multi-directional sensemaking throughout the process. Our data shows that the cycles between activities were tightly linked. For example, participants used incoming data to gain an initial understanding of a model's capabilities. The complex process of data collection, selection, and pre-processing is tightly linked and often involves multiple instances of interpretation. Many companies have historical data collected without specific application or consistent strategy in mind. This data may change with new employees or the introduction of AI into the data-producing processes. Some MLOps models suggest data exploration but provide limited guidance on how it should be conducted, often treating data as a measurable artefact [49]. They also overlook the necessary infrastructure for appropriate data access [52]. We argue that an iterative interpretation process around data and its infrastructure is crucial for subsequent value generation, ensuring the system is used correctly to produce relevant insights.

Our study further supports the critique of MLOps in an interorganizational context [52]. Interviews revealed differing practices, experiences, and expectations among organizations. Notably, interpretations of "development" varied. Clients assumed it included data-related activities managed by IBM, while IBM's development services were more narrowly focused on data analysis, model training, and application construction. This discrepancy caused significant issues. Similarly, interpretations of concepts like *data*, *product*, *Watson expert*, *training*, *testing*, and *project* were negotiated mid-project, highlighting the importance of clear communication and shared understanding in MLOps projects. Our study revealed that spontaneous distribution of responsibilities, such as *ad hoc* data creation, often led to project disruptions. This shifting of roles and responsibilities was often due to unforeseen demands from technology, data, or context. Previous SE4AI research often discussed the roles of developers [4], [16] or data scientists [23], [28], but we found that clients, even in roles like product owner, are significantly affected by these AI-related challenges. MLOps could potentially address these issues as it outlines various roles and responsibilities [48]. However, it is unclear how these roles should be divided between partners in an interorganizational setting and how role interfaces should be defined across organizations.

SE4AI literature often isolates model training as the central development step, separating it from production and delivery [9], [44], [45], [48], [49], [53]. However, modern AIDPs provide pre-set pipelines and configurations that bridge this gap. A prime example is Q&A chatbot applications, where the tools automatically generate a chatbot instance based on data sets or question-answer pairs. While this technology responds to users' or clients' desire for a quick, tangible systems over statistical accuracy assessments, it can complicate the evaluation of individual components. For instance, it may be hard to discern if quality issues stem from the data or the Q&A model. This opaqueness is increasing with the use of general-purpose pre-trained LLMs. Developers need specific guidance to balance the need for interpretability and the desire for early tangible results. We also advocate for research acknowledging AIDPs and pre-trained models' potential role in AI delivery, e.g., guidance on assessing ways to achieve functionality, such as deciding whether to use out-of-the-box functionality, fine-tuning, or training a new model.

Considering these insights, we propose that a comprehensive approach to deployment and operation of AI capabilities must account for the following activities:

1. Continuous *sensemaking of AIDPs and models*, and *how they can be used in the projected use case*. This outcome can be achieved by benchmarking similar cases, testing models with sample data, comparing costs and runtimes, and leveraging the expertise of other users or providers. All team members should participate in tests with real or test data to collectively comprehend the technology.

2. Continuous *sensemaking of available data* and *data infrastructure*, e.g., investigating data sources and data creation processes, tracking dataset changes, and understanding its current usage. Further, one can test data samples with various AI tools, models, and AIDPs to gain AI-driven insights and uncover unknown unknowns.

3. Regular *sensemaking of context, requirements, and expectations* using, when possible, tangible prototypes. Rather than using best-case solutions or demos, one should allow domain experts and users to understand the technology through prototype interaction. These prototypes, based on simple models, can be updated or fine-tuned later. As interviews suggest, people are adept at evaluating interfaces from visual cues but struggle with non-visual systems.

4. Regular *sensemaking of the project structure, roles, progress, and goals*. Initially, prior experiences or standard development models should be made explicit to make expectations about the partner's role obvious. Later, teams should revisit these initial statements to discuss any changes or shifts. Project members should value learning as a success metric and reward sensemaking and learning processes that contribute to the studied use case. They need to document key learnings, especially those highlighting technology and data limitations. As AI technology evolves rapidly, current challenges may become easier to solve based on documented learnings, providing organizations with an advantage in making progress. Accordingly, project managers need to allocate resources to sensemaking. They should reconsider roles and responsibilities based on specific knowledge and sensemaking capabilities, e.g., by assigning some tasks related to data to persons who can best make sense of them, rather than following an outsourcing schema. Finally, they need to embrace planning uncertainty as element in AIDP.

We propose *heedful interrelating* as the vision for an adequate project collaboration. Heedful interrelating characterizes the nature of social relationships that support collectives at achieving their goals [85]. A potential operationalization of heedful interrelating involves contributing and taking actions to support others, subordinating one's own actions to fit with the actions of others, and, most importantly, "envisioning the system of collective work being realized by the team as a whole" [86, p. 2]. This perspective implies that project members consider what interpretations and meanings might emerge based on their own actions, and how those actions as well as resulting meanings will impact the project structure, roles, or goals. This consideration demands a mindful approach towards one's own and other's actions.

To enable mindfulness, we suggest that project members and managers embrace exploration and learning as a way to make progress. Participants should engage early in identifying potential targets for sensemaking and watching out for breakdowns, which implies asking oneself and others if actions or events occurred in an unexpected manner. This heedfulness demands a culture in which admitting that something goes against plan is rewarded rather than considered a sign of ineffectiveness. Finally, project members should plan for interactions between sensemaking targets as denoted by the concept of *waves of sensemaking*. If breakdown occurred in one area, it is possible that it will necessitate sensemaking in another area as well.

Overall, software engineers and project managers will benefit from adopting mindfulness as an approach to leading and conducting AIDP projects. We claim that equipped with the above guidance, they will be able to spot upcoming problems and challenges with greater ease. While this is likely to hold for all SE projects, we see it as particularly important in projects involving AI, as it produces unexplainable, contradictory results more often than legacy technologies, which then might impact on projects structure. Also, careful sensemaking about the larger context is necessary to adequately react to themes appearing in public discourse. AI has become a major topic among general public, which might introduce preconceptions that need to be addressed during design and development, as well as when deploying the solution. Finally, the inherent complexity of AI related to its non-deterministic nature will necessitate more frequent sensemaking. Awareness of it might be necessary to assess the (lack of) progress in the project accordingly.

### Threats to validity

This study has several limitations due to its scope, time frame, location, methodology, and theoretical perspective. This study, focused on IBM Watson projects, may have limitations concerning external validity due to rapid AI advancements and the rise of LLMs. The acquisition of experts may now extend beyond the provider's direct workforce. For instance, OpenAI fosters an independent developer community, unlike IBM, which primarily profits from consulting and expertise. However, the general challenge of expert access persists, potentially increasing need for sensemaking as clients decide on expertise sources (IT consultancy, freelancer agencies, own resources) and skill assessments. Given this study's limited scope, further research on sensemaking in projects using other AIDPs and recent AI paradigms is recommended.

Our study focused on collaborations between Swiss companies and the local IBM branch in 2017. This specific temporal and geographical context, as well as potential local sales strategies, may have shaped client expectations.

To ensure external validity, it's recommended to replicate the results in different contexts, perhaps through a wider survey. The limited number of cases and our decision to interview both company and IBM representatives per case may introduce bias, as some companies with confidential projects may have opted out. Additional incentives could encourage such companies to participate. Long-term study or ethnography could also provide more comprehensive insights into ongoing sensemaking processes. Further, we have focused on sensemaking in the software construction phase. Important sensemaking occur in the post-deployment phase, which require dedicated research efforts. For instance, prior work has described problems due to data drift during use, which our snapshot did not encounter.

Our study's reliance on qualitative data, specifically interviews, carries inherent risks. It is based on retrospective analysis, which may be subject to availability bias and reflect an individual's perception of past events. However, this approach did enable participants to share their insights and reflections, revealing their sensemaking efforts. Ethnographic studies could further validate these findings. A survey could provide quantified results, e.g., about the frequency of specific challenges or episodes.

Our study's theoretical lens, focusing on sensemaking, may have highlighted certain elements while downplaying others. This choice was guided by initial data findings, not pre-determined. Future studies could benefit from a multi-theory approach using theories about education, outsourcing, or collaboration. This would provide a more comprehensive view and multidimensional interpretation of relevant incidents. Despite its limitations, our study can stimulate further research in SE and other fields.

## Conclusion and Implications for Practice

The collected data indicates that the complexity of AIDP-based development is increased compared to traditional systems development. Uncertainty about what is possible now or what will be possible in the future and the dependence of technology capabilities on the specific data and input make the situation harder to grasp and enact.

Our study underscores the importance of embracing learning in SE practice. Recognizing data limitations and algorithm constraints is crucial for successful AI-based development projects. Traditional quantitative accuracy metrics may inadequately gauge progress, potentially leading to premature project termination.

Developers and clients should be vigilant for breakdowns of meanings in AI-based projects, given their novelty and rapid technological advancements. They should assume their understandings tentative, more so than in other situations. AI's probabilistic output may occasionally surprise project members, necessitating more sensemaking around technology and data than in deterministic projects. This, combined with the *waves of sensemaking*, suggests that AI projects may face more disruptions and recoveries, requiring balanced mindfulness. Project members should carefully identify potential sensemaking targets and anticipate breakdowns. They should also plan for interferences in sensemaking across all areas, managing it as a learning opportunity rather than a hindrance.

Informants who observed the heightened need for sensemaking responded by enhancing collaboration and communication. This interaction resulted in continuous input and feedback loops between partners, intensifying efforts regardless of the participants' organizational identity.

The study uses data collected in Switzerland and describes projects between Swiss companies and an international AIDP provider. This adds to previous studies which were conducted in Asia [4] or North America [16], [23], theorized based on literature review [9], [24], or attended to a single group of professionals like developers or data scientists [16], [23], [28]. The results offer a socio-cognitive viewpoint to complement other SE4AI studies, aligning with the qualitative research ideals of concatenation and cumulation [87], [88]. The employed method provides a deep and multi-sided perspective on the AIDP-based development allowing for exploration of sensemaking processes.

## Acknowledgements

We express our best gratitude to Daniel Oettli and Nicola Storz for their engagement in collecting the data which form the basis for this manuscript. We also thank Dr. Alain Gut and Philip Spaeti from IBM Switzerland for their fundamental support throughout the research project. Finally, our gratitude goes to Gerhard Schwabe for enabling this collaboration and providing his invaluable guidance.

## References

[1] C. Collins, D. Dennehy, K. Conboy, and P. Mikalef, "Artificial intelligence in information systems research: A systematic literature review and research agenda," *International Journal of Information Management*, vol. 60, p. 102383, Oct. 2021,

[2] T. Davenport and R. Kalakota, "The potential for artificial intelligence in healthcare," *Future Healthc J*, vol. 6, no. 2, pp. 94–98, Jun. 2019,

[3] M. Dolata and G. Schwabe, "How Fair Is IS Research?," in *Engineering the Transformation of the Enterprise: A Design Science Research Perspective*, S. Aier, P. Rohner, and J. Schelp, Eds. Cham: Springer International Publishing, 2021, pp. 37–49.

[4] Z. Wan, X. Xia, D. Lo, and G. C. Murphy, "How does Machine Learning Change Software Development Practices?," *IEEE Trans on Software Engineering*, vol. 47, no. 9, pp. 1857–1871, Sep. 2021,

[5] M. Dolata, S. Feuerriegel, and G. Schwabe, "A sociotechnical view of algorithmic fairness," *Information Systems Journal*, vol. 32, no. 4, pp. 754–818, 2022,

[6] NSTC, "The National Artificial Intelligence Research and Development Strategic Plan: 2019 Update," Select Committee on Artificial Intelligence of the National Science & Technology Council, Washington, D.C, Technical Report AD1079707, 2019. Available: https://apps.dtic.mil/sti/citations/AD1079707

[7] M. Mitchell, *Artificial intelligence: a guide for thinking humans*. 2019.

[8] M. Minsky, "Logical vs. analogical or symbolic vs. connectionist or neat vs. scruffy," in *Artificial intelligence at MIT expanding frontiers*, Cambridge, MA, USA: MIT Press, 1991, pp. 218–243.

[9] S. Martínez-Fernández *et al.*, "Software Engineering for AI-Based Systems: A Survey," *ACM Trans. Softw. Eng. Methodol.*, vol. 31, no. 2, p. 37e:1-37e:59, Apr. 2022,

[10] E. Nascimento, A. Nguyen-Duc, I. Sundbø, and T. Conte, "Software engineering for artificial intelligence and machine learning software: A systematic literature review." arXiv, Nov. 07, 2020. Available: http://arxiv.org/abs/2011.03751

[11] S. E. Page, "What sociologists should know about complexity," *Annual Review of Sociology*, vol. 41, pp. 21–41, 2015.

[12] M. Mitchell, *Complexity: a guided tour*. Oxford [England]; New York: Oxford University Press, 2009.

[13] K. E. Weick and K. M. Sutcliffe, *Managing the unexpected: sustained performance in a complex world*, Third edition. Hoboken, New Jersey: John Wiley & Sons, Inc, 2015.

[14] D. Riehle, M. Capraro, D. Kips, and L. Horn, "Inner Source in Platform-Based Product Engineering," *IEEE Transactions on Software Engineering*, vol. 42, no. 12, pp. 1162–1177, Dec. 2016,

[15] L. Prechelt, "Plat_Forms: A Web Development Platform Comparison by an Exploratory Experiment Searching for Emergent Platform Properties," *IEEE Transactions on Software Engineering*, vol. 37, no. 1, pp. 95–108, Jan. 2011,

[16] C. T. Wolf and D. Paine, "Sensemaking Practices in the Everyday Work of AI/ML Software Engineering," in *Proceedings of the IEEE/ACM 42nd International Conference on Software Engineering Workshops*, New York, NY, USA, Sep. 2020, pp. 86–92.

[17] A. Gawer, Ed., *Platforms, markets, and innovation*. Cheltenham, UK ; Northampton, MA: Edward Elgar, 2009.

[18] M. Dolata, K. Crowston, and G. Schwabe, "Project Archetypes: A Blessing and a Curse for AI Development," in *Proc. Intl. Conf. on Information Systems*, Dec. 2022.

[19] J. M. Tarn, D. C. Yen, and M. Beaumont, "Exploring the rationales for ERP and SCM integration," *Industrial Management & Data Systems*, vol. 102, no. 1, pp. 26–34, Jan. 2002,

[20] R. Malhotra and C. Temponi, "Critical decisions for ERP integration: Small business issues," *International Journal of Information Management*, vol. 30, no. 1, pp. 28–37, Feb. 2010,

[21] L. Kharb, "A Perspective View on Commercialization of Cognitive Computing," in *Intl. Conf. on Cloud Computing, Data Science Engineering (Confluence)*, Jan. 2018, pp. 829–832.

[22] G. Giray, "A software engineering perspective on engineering machine learning systems: State of the art and challenges," *Journal of Systems and Software*, vol. 180, p. 111031, Oct. 2021,

[23] M. Kim, T. Zimmermann, R. DeLine, and A. Begel, "Data Scientists in Software Teams: State of the Art and Challenges," *IEEE Transactions on Software Engineering*, vol. 44, no. 11, pp. 1024–1038, Nov. 2018,

[24] F. Khomh, B. Adams, J. Cheng, M. Fokaefs, and G. Antoniol, "Software Engineering for Machine-Learning Applications: The Road Ahead," *IEEE Software*, vol. 35, no. 5, pp. 81–84, Sep. 2018,

[25] S. Maitlis and S. Sonenshein, "Sensemaking in Crisis and Change: Inspiration and Insights From Weick (1988)," *Journal of Management Studies*, vol. 47, no. 3, pp. 551–580, 2010,

[26] K. E. Weick, *Sensemaking in organizations*. Thousand Oaks: Sage Publications, 1995.

[27] K. E. Weick, "The Collapse of Sensemaking in Organizations: The Mann Gulch Disaster," *Administrative Science Quarterly*, vol. 38, no. 4, pp. 628–652, 1993,

[28] Á. A. Cabrera, M. T. Ribeiro, B. Lee, R. DeLine, A. Perer, and S. M. Drucker, "What Did My AI Learn? How Data Scientists Make Sense of Model Behavior," *ACM Trans. Comput.-Hum. Interact.*, Jun. 2022.

[29] M. Hesenius, N. Schwenzfeier, O. Meyer, W. Koop, and V. Gruhn, "Towards a Software Engineering Process for Developing Data-Driven Applications," in *2019 IEEE/ACM 7th International Workshop on Realizing Artificial Intelligence Synergies in Software Engineering (RAISE)*, May 2019, pp. 35–41.

[30] A. Arpteg, B. Brinne, L. Crnkovic-Friis, and J. Bosch, "Software Engineering Challenges of Deep Learning," in *2018 44th Euromicro Conference on Software Engineering and Advanced Applications (SEAA)*, Aug. 2018, pp. 50–59.

[31] G. Lorenzoni, P. Alencar, N. Nascimento, and D. Cowan, "Machine Learning Model Development from a Software Engineering Perspective: A Systematic Literature Review." arXiv, Feb. 15, 2021. Available: http://arxiv.org/abs/2102.07574

[32] P. Bourque and R. E. Fairley, *Guide to the Software Engineering Body of Knowledge - SWEBOK V3.0*. 2014.

[33] A. Paleyes, R.-G. Urma, and N. D. Lawrence, "Challenges in Deploying Machine Learning: A Survey of Case Studies," *ACM Comput. Surv.*, vol. 55, no. 6, p. 114:1-114:29, Dezember 2022,

[34] N. Nahar, H. Zhang, G. Lewis, S. Zhou, and C. Kästner, "A Meta-Summary of Challenges in Building Products with ML Components -- Collecting Experiences from 4758+ Practitioners," in *Proc. Intl. Conf. on AI Engineering - Software Engineering for AI*, Melbourne, Australia, Mar. 2023.

[35] S. Huang, E.-H. Liu, Z.-W. Hui, S.-Q. Tang, and and S.-J. Zhang, "Challenges of Testing Machine Learning Applications," *Intl. Journal of Performability Engineering*, vol. 14, no. 6, p. 1275, 2018,

[36] J. Gao, C. Tao, D. Jie, and S. Lu, "Invited Paper: What is AI Software Testing? and Why," in *IEEE Intl. Conference on Service-Oriented System Engineering (SOSE)*, Apr. 2019, pp. 27–2709.

[37] P. Ralph, "Software engineering process theory: A multi-method comparison of Sensemaking–Coevolution–Implementation Theory and Function–Behavior–Structure Theory," *Information and Software Technology*, vol. 70, pp. 232–250, Feb. 2016,

[38] X. Franch, A. Jedlitschka, and S. Martínez-Fernández, "A Requirements Engineering Perspective to AI-Based Systems Development: A Vision Paper," in *Requirements Engineering: Foundation for Software Quality*, Cham, 2023, pp. 223–232.

[39] L. E. G. Martins and T. Gorschek, "Requirements Engineering for Safety-Critical Systems: An Interview Study with Industry Practitioners," *IEEE Trans. Software Engineering*, vol. 46, no. 4, pp. 346–361, Apr. 2020,

[40] Z. Pei, L. Liu, C. Wang, and J. Wang, "Requirements Engineering for Machine Learning: A Review and Reflection," in *2022 IEEE 30th International Requirements Engineering Conference Workshops (REW)*, Aug. 2022, pp. 166–175.

[41] K. Ahmad, M. Abdelrazek, C. Arora, M. Bano, and J. Grundy, "Requirements engineering for artificial intelligence systems: A systematic mapping study," *Information and Software Technology*, vol. 158, p. 107176, Jun. 2023,

[42] K. Ahmad, M. Abdelrazek, C. Arora, A. Agrahari Baniya, M. Bano, and J. Grundy, "Requirements engineering framework for human-centered artificial intelligence software systems," *Applied Soft Computing*, vol. 143, p. 110455, Aug. 2023,

[43] G. Hains, A. Jakobsson, and Y. Khmelevsky, "Towards formal methods and software engineering for deep learning: Security, safety and productivity for dl systems development," in *2018 Annual IEEE Intl. Systems Conf. (SysCon)*, Apr. 2018, pp. 1–5.

[44] L. E. Lwakatare, A. Raj, J. Bosch, H. H. Olsson, and I. Crnkovic, "A Taxonomy of Software Engineering Challenges for Machine Learning Systems: An Empirical Investigation," in *Agile Processes in Software Engineering and Extreme Programming*, Cham, 2019, pp. 227–243.

[45] T. Menzies, "The Five Laws of SE for AI," *IEEE Software*, vol. 37, no. 1, pp. 81–85, Jan. 2020,

[46] M. Steidl, M. Felderer, and R. Ramler, "The pipeline for the continuous development of artificial intelligence models—Current state of research and practice," *Journal of Systems and Software*, vol. 199, p. 111615, May 2023,

[47] K. Shivashankar and A. Martini, "Maintainability Challenges in ML: A Systematic Literature Review," in *Euromicro Conf- on Software Engineering and Advanced Applications (SEAA)*, Aug. 2022, pp. 60–67.

[48] M. M. John, H. H. Olsson, and J. Bosch, "Towards MLOps: A Framework and Maturity Model," in *2021 47th Euromicro Conference on Software Engineering and Advanced Applications (SEAA)*, Sep. 2021, pp. 1–8.

[49] M. Treveil, *Introducing MLOps: how to scale machine learning in the enterprise*, First edition. Beijing ; Boston: O'Reilly, 2020.

[50] A. Bhatia, E. E. Eghan, M. Grichi, W. G. Cavanagh, Z. M. (Jack) Jiang, and B. Adams, "Towards a change taxonomy for machine learning pipelines," *Empir Software Eng*, vol. 28, no. 3, p. 60, 2023,

[51] S. Amershi *et al.*, "Software Engineering for Machine Learning: A Case Study," May 2019, pp. 291–300.

[52] T. Granlund, A. Kopponen, V. Stirbu, L. Myllyaho, and T. Mikkonen, "MLOps Challenges in Multi-Organization Setup: Experiences from Two Real-World Cases," in *2021 IEEE/ACM 1st Workshop on AI Engineering - Software Engineering for AI (WAIN)*, May 2021, pp. 82–88.

[53] X. Zhang, Y. Yang, Y. Feng, and Z. Chen, "Software Engineering Practice in the Development of Deep Learning Applications." arXiv, Oct. 07, 2019. Available: http://arxiv.org/abs/1910.03156

[54] I. D. Raji *et al.*, "Closing the AI accountability gap: defining an end-to-end framework for internal algorithmic auditing," in *Proc. Conf. on Fairness, Accountability, and Transparency*, New York, NY, USA, Jan. 2020, pp. 33–44.

[55] S. Schelter, F. Biessmann, T. Januschowski, D. Salinas, S. Seufert, and G. Szarvas, "On Challenges in Machine Learning Model Management," *Bulletin of the IEEE Computer Society Technical Committee on Data Engineering*, pp. 5–13, 2018.

[56] M. Arnold *et al.*, "FactSheets: Increasing Trust in AI Services through Supplier's Declarations of Conformity." arXiv, Feb. 07, 2019. Available: http://arxiv.org/abs/1808.07261

[57] N. Naik, "Crowdsourcing, Open-Sourcing, Outsourcing and Insourcing Software Development: A Comparative Analysis," in *2016 IEEE Symposium on Service-Oriented System Engineering (SOSE)*, Mar. 2016, pp. 380–385.

[58] D. Èmite and C. Wohlin, "A Whisper of Evidence in Global Software Engineering," *IEEE Software*, vol. 28, no. 4, pp. 15–18, Jul. 2011,

[59] R. Dzhusupova, J. Bosch, and H. H. Olsson, "The goldilocks framework: towards selecting the optimal approach to conducting AI projects," in *Proc. Intl. Conf. on AI Engineering: Software Engineering for AI*, New York, NY, USA, Oktober 2022, pp. 124–135.

[60] P. Ralph, "Toward Methodological Guidelines for Process Theories and Taxonomies in Software Engineering," *IEEE Transactions on Software Engineering*, vol. 45, no. 7, pp. 712–735, Jul. 2019,
[61] P. W. L. Vlaar, P. C. van Fenema, and V. Tiwari, "Cocreating Understanding and Value in Distributed Work: How Members of Onsite and Offshore Vendor Teams Give, Make, Demand, and Break Sense," *MIS Quarterly*, vol. 32, no. 2, p. 227, 2008,
[62] R. Holt and J. Cornelissen, "Sensemaking revisited," *Management Learning*, vol. 45, no. 5, pp. 525–539, Nov. 2014,
[63] T. B. Jensen, A. Kjærgaard, and P. Svejvig, "Using Institutional Theory with Sensemaking Theory: A Case Study of Information System Implementation in Healthcare," *Journal of Information Technology*, vol. 24, no. 4, pp. 343–353, Dec. 2009,
[64] K. E. Weick, K. M. Sutcliffe, and D. Obstfeld, "Organizing and the Process of Sensemaking," *Organization Science*, vol. 16, no. 4, pp. 409–421, Aug. 2005,
[65] R. Gregory, "Political Responsibility for Bureaucratic Incompetence: Tragedy at Cave Creek," *Crisis management*, vol. 3, no. 1, pp. 269–285, 2008.
[66] K. E. Weick, K. M. Sutcliffe, and D. Obstfeld, "Organizing for high reliability: Processes of collective mindfulness," *Crisis management*, vol. 3, no. 1, pp. 81–123, 2008.
[67] J. McAvoy, T. Nagle, and D. Sammon, "Using mindfulness to examine ISD agility," *Information Systems Journal*, vol. 23, no. 2, pp. 155–172, 2013,
[68] B. A. Turner, *Man-made Disasters*. Wykeham, 1978.
[69] K. E. Weick, "Enacted Sensemaking in Crisis Situations," *Journal of Management Studies*, vol. 25, no. 4, pp. 305–317, 1988,
[70] I. Stigliani and D. Ravasi, "Organizing Thoughts and Connecting Brains: Material Practices and the Transition from Individual to Group-Level Prospective Sensemaking," *Academy of Management Journal*, vol. 55, no. 5, pp. 1232–1259, Oct. 2012,
[71] K. E. Weick, "Sensemaking as an Organizational Dimension of Global Change," in *Organizational Dimensions of Global Change: No Limits to Cooperation*, Thousand Oaks, CA: SAGE Publications, Inc., 1999, pp. 39–56.
[72] K. Hwang, *Cloud Computing for Machine Learning and Cognitive Applications*. Cambridge, Massachusetts: The MIT Press, 2017.
[73] J. E. Kelly and S. Hamm, *Smart Machines: IBM's Watson and the Era of Cognitive Computing*. New York: Columbia University Press, 2013.
[74] R. A. Stebbins, *Exploratory research in the social sciences*. Thousand Oaks, Calif.: Sage Publications, 2001.
[75] J. C. Flanagan, "The critical incident technique.," *Psychological bulletin*, vol. 51, no. 4, p. 327, 1954.
[76] D. Keatinge, "Versatility and flexibility: Attributes of the Critical Incident Technique in nursing research," *Nursing & Health Sciences*, vol. 4, no. 1–2, pp. 33–39, Mar. 2002,
[77] R. K. Yin, "Applications of case study research," *Series, 4th. Thousand Oaks: Sage Publications*, 2003.
[78] P. Runeson and M. Höst, "Guidelines for conducting and reporting case study research in software engineering," *Empirical software engineering*, vol. 14, no. 2, p. 131, 2009.
[79] D. Staehelin, M. Dolata, and G. Schwabe, "Managing Tensions in Research Consortia with Design Thinking Artifacts," in *Design Thinking for Software Engineering: Creating Human-oriented Software-intensive Products and Services*, J. Hehn, D. Mendez, W. Brenner, and M. Broy, Eds. Cham: Springer Intl., 2022, pp. 137–153.
[80] P. Ralph, "The Sensemaking-Coevolution-Implementation Theory of software design," *Science of Computer Programming*, vol. 101, pp. 21–41, Apr. 2015,
[81] S. Faily *et al.*, "Requirements Sensemaking Using Concept Maps," in *Human-Centered Software Engineering*, Berlin, Heidelberg, 2012, pp. 217–232.
[82] B. Shreeve, P. Ralph, P. Sawyer, and P. Stacey, "Geographically distributed sensemaking: developing understanding in forum-based software development teams," in *Proc. Intl. Workshop on Cooperative and Human Aspects of Software Engineering*, Florence, Italy, Mai 2015, pp. 36–42.
[83] Z. Wang, Y. Feng, Y. Wang, J. A. Jones, and D. Redmiles, "Unveiling Elite Developers' Activities in Open Source Projects," *ACM Trans. Softw. Eng. Methodol.*, vol. 29, no. 3, p. 16:1-35, 2020,
[84] M. Aanestad and T. B. Jensen, "Collective Mindfulness in Post-implementation IS Adaptation Processes," *Accounting Management and Information Technologies*, vol. 26, no. 1–2, pp. 13–27, 2016,
[85] K. E. Weick and K. H. Roberts, "Collective Mind in Organizations: Heedful Interrelating on Flight Decks," *Administrative Science Quarterly*, vol. 38, no. 3, p. 357, Sep. 1993,
[86] J. P. Stephens and C. J. Lyddy, "Operationalizing Heedful Interrelating: How Attending, Responding, and Feeling Comprise Coordinating and Predict Performance in Self-Managing Teams," *Frontiers in Psychology*, vol. 7, 2016.
[87] R. A. Stebbins, "Concatenated Exploration: Aiding Theoretic Memory by Planning Well for the Future," *Journal of Contemporary Ethnography*, vol. 35, no. 5, pp. 483–494, Oct. 2006,
[88] P. Dourish, "Reading and Interpreting Ethnography," in *Ways of Knowing in HCI*, J. S. Olson and W. A. Kellogg, Eds. New York, NY: Springer, 2014, pp. 1–23.

**Mateusz Dolata** is a Researcher at the Department of Informatics, University of Zurich. He earned his Ph.D. in 2018, focusing on the digital transformation within advisory services. He studies the use of computer-supported collaboration and artificial intelligence (AI) for societal benefit. His research encompasses the employment of AI-based agents in education, healthcare, emergency response, and customer service. Also, he explores the societal implications of emerging technological developments such as AI, metaverse, and drones contributing a new perspective on the sociotechnical aspects of those advancements. He receives financial support from SNSF and INNOSUISSE.

**Kevin Crowston** is a Distinguished Professor of Information Science in the School of Information Studies at Syracuse University. He received his Ph.D. (1991) in Information Technologies from the Sloan School of Management, Massachusetts Institute of Technology (MIT). His research examines new ways of organizing made possible by the extensive use of information and communications technology. Specific research topics include the development practices of Free/Libre Open Source Software teams and work practices and technology support for citizen science research projects, both with NSF support. His most recent project is a study of the evolution of newswork with new technologies.

# Supplemental material

This is a supplement for the manuscript "Making sense of AI systems development" by Mateusz Dolata and Kevin Crowston, IEEE Transactions on Software Engineering, 2023.

Table A1. Overview of the projects considered in the study. Interview codes with C in the middle indicate client-side employees, the one with V—IBM-side employees.

| No. | Firm | Interviewees | Sector | Short Description |
|---|---|---|---|---|
| 1 | A | 1C1, 1V1 | Telecommunications | Automated email categorization for internal and incoming email exchange |
| 2 | B | 2V1, 2C1 | Retail Banking | Enterprise search for unstructured data (e.g,. textual process descriptions) |
| 3 | C | 3C1, 3V1, 3V2 | Retail Banking | Know your supplier - 360° view of information about potential suppliers |
| 4 | C | 3C1, 3V1, 3V2 | Retail Banking | Deliverables and Obligations - analysing supply contract data |
| 5 | D | 5C1, 5V1, 5V2 | Transportation | Chatbot/robot for guiding passengers at a transportation hub |
| 6 | E | 6V1 | Consumer products | Regulatory and legal information search and summary about specific ingredients |
| 7 | F | 7C1, 7V1 | IT services | Chatbot for password request if a user needs new password |
| 8 | G | 8C1, 8V1 | Insurance | Make insurance claims searchable for comparison when new claims arrive |
| 9 | G | 8C1, 8V1 | Insurance | Make insurance contracts in health and life insurances searchable for similar cases |
| 10 | H | 10V1 | Healthcare | Initial patient allocation to doctors and stations / confidence-induced reasoning |
| 11 | I | 11C1, 11V1 | Machine industry | Collaborative robot for Industry 4.0 using sensory inputs and voice commands |
| 12 | J | 12C1, 12V1 | Insurance | Underwriting (checking risks level) for small business customers |
| 13 | K | 13C1, 13V1 | Chemical industry | Single point of access for company data distributed across platforms / 360° view |
| 14 | L | 14C1, 14V1 | Machine industry | Customer feedback form (open text) analysis / identification of complaints |
| 15 | L | 14C1, 14V1 | Machine industry | Sales Tenders: Analysis of Calls for Proposals to assess whether should bid |
| 16 | L | 14C1, 14V1 | Machine industry | Finance Dept.: Predicting course of business of suppliers and customers |
| 17 | G | 17C1, 17V1 | Insurance | Service Desk Integration and information provision for help desk agents |
| 18 | M | 18C1, 18V1 | IT services | Enterprise search for unstructured data for IT management and CRM |
| 19 | N | 19V1, 19V2 | Consumer products | Company-specific chatbot-based service desk for ERP-related issues |
| 20 | O | 20V1 | Chemical industry | Analysis of potentially relevant or successful startups for investment purposes |
| 21 | P | 21C1, 21V1, 21V2 | Public sector | Filtering out noise and identifying potential clues relevant in investigative work |

Table A2. Summary of the coding schema used in the first round of coding.
Meta-codes predefined; sub-codes emerged bottom-up.

| Meta-Code | #Segments | Short Description and Most Prominent Sub-Codes |
|---|---|---|
| Application Domain | 799 | Focuses on reasons for and ways of employing Watson by companies: Use Case, Goals, Reasons for choosing Watson, Influence on Competition, Potential Future Use Cases |
| Project Management | 1270 | Focuses on typical characteristics of Watson-based projects and their implications: Use Case Elaboration, System Deployment and Integration, Model Training, Collaboration, Evaluation, Management Approach, General Characteristics |
| Requirements for IBM Watson | 1135 | Focuses on what is necessary for a company to successfully develop and deploy Watson-based apps: Human Resources, Project Mindset/Culture, Data, Strategy, Other Requirements |
| Impact on Individual/Human-Computer Interaction | 720 | Focuses on how future users and their work will be impacted by Watson-based applications: Change Management, Process Shift, Role of Watson, Role of Human, Replacement of Human Being, Future Skills |

The coding schema used in the second round was driven by the process of sensemaking (sources of initial expectations, sources of breakdown, new provisional meanings) and codes are presented in the tables provided in Findings.

Table A3. Steps and strategies applied by the research team to assure the reliability and validity of the research presented along the dimensions proposed by Merriam and Grenier, 2019, *Qualitative research in practice: examples for discussion and analysis*, second edition.

| Strategy | Implementation |
|---|---|
| Triangulation | The analysis uses a set of 36 interviews about 21 different projects from collaboration between IBM and 17 different companies. Additionally, the data was collected by two different researchers in parallel and then coded in two rounds: first, by them, and second, by a researcher experienced in qualitative data analysis. |
| Member checks | The interviews were provided to the interviewees for final proof. The results of the initial interpretations were discussed with IBM representatives in workshops. Later results were provided to IBM representatives in writing for feedback. |
| Peer review/ examination | The obtained results were regularly consulted with an additional senior researcher who was involved in setting up the collaboration and in the supervision of the student assistants. He as well as one external reviewer who has the same affiliation as the second author provided extended feedback to an earlier draft of this paper. |
| Researcher's position or reflexivity | Neither authors nor any person involved in the project (student assistants, scientific personnel) received funding from IBM or any of the companies participating in the study. The contact person who initiated the collaboration on IBM's side and participated in project selection as well as interpretation interviews is an alumnus of the first author's University and this is how the collaboration emerged. The researchers involved in the project were self-funded by their respective organizations. The primary motivation to engage in this research was to learn more about development involving AI and share it with the research community. The secondary motivation was to help IBM understand challenges, especially managerial challenges, occurring in the Watson projects.<br>The interviewers were two Master-level students of Informatics. Both were Swiss and had experience in development of software and in innovation projects. Both worked at external, IT-related organizations in parallel to studying. They did not know the interviewees before the study, nor did they have direct connection to IBM or IBM Watson before. They embraced the perspective of researchers who want to explore a new type of projects and a new technology. The students were overseen by qualitative researchers. Notes from interviewees were regularly reviewed resulting in further guidance for subsequent rounds of interviews. |
| Adequate engagement in data collection | Both authors were engaged in the supervision of the students who collected the interviews. This involves establishing contact between them and IBM, preparation of the interview guide, training of the student assistants in techniques for conducting interviews, and suggestions on what might be specifically interesting to discuss for various projects. They also supported the initial round of coding via feedback and provision of methodological knowledge. Further, they were directly involved in the workshops conducted between the University and IBM. |
| Maximum variation | The variation is limited by the fact that the study focuses on IBM Watson, the primary interview partners were in leading positions in their respective projects, and that the projects were pre-selected by IBM. However, no systematic elicitation of demographic data in this regard was conducted as it was not at the core of the study. Particularly, we do not make statements based on demographic differences or similarities. However, based on the interviewee's subjective appraisal, the interviewees were very diverse in terms of background, origin, experience, age, or status within their company. |
| Audit trail | The selection and analysis of the data were documented and controlled, thus ensuring that the process and decisions taken along the process are reproducible. This manuscript serves as a display of the analysis process and exemplary results. |
| Rich, thick descriptions | The manuscript provides an in-depth exploration of the data collected and analysed by the researchers, with many extensive excerpts from the data demonstrating the findings and their implications. This allows the reader to assess the conclusions drawn from the data and their implications. |

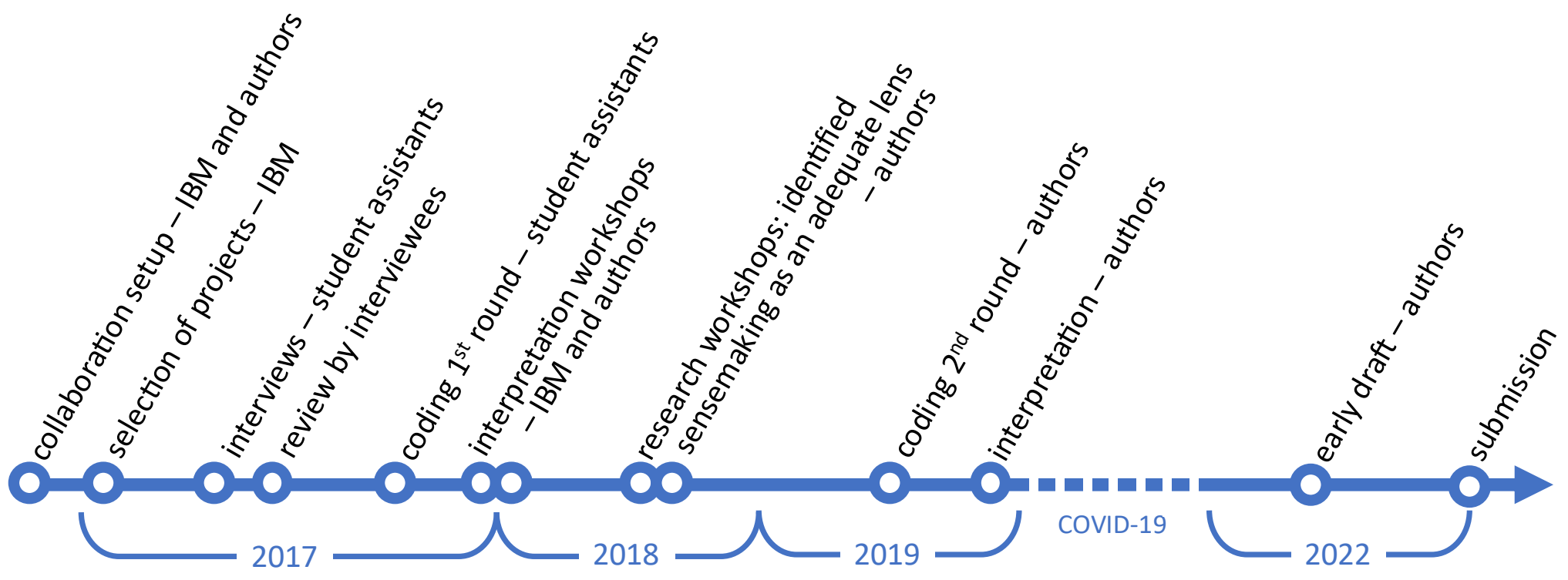

Figure A1. The timeline of the research project with the indication of responsibilities.